\documentclass[aps,prd,reprint,twocolumn,superscriptaddress,nofootinbib,floatfix,longbibliography,preprintnumbers]{revtex4-2}
\usepackage[a4paper, margin=2.0cm]{geometry}
\usepackage{booktabs}   % \toprule \midrule \bottomrule
\usepackage{siunitx}    % S columns + \si
\usepackage{tabularx}   % tabularx environment
\usepackage{array}      % >{\raggedright\arraybackslash} and \arraybackslash
\usepackage{amsmath,amssymb,mathtools,bm,braket,mathrsfs}
\DeclareSIUnit{\parsec}{pc}
\DeclareSIUnit{\decihertz}{dHz}
\DeclareSIUnit{\Msun}{\text{M}_{\odot}}
\usepackage{graphicx}
\usepackage[caption=false]{subfig}
\graphicspath{{./plot/}}
\usepackage{booktabs,multirow}
\usepackage{dcolumn} % decimal aligned columns: D{.}{.}{-1}
\usepackage[final]{microtype}
\usepackage[dvipsnames]{xcolor}
\definecolor{refblue1}{RGB}{0, 160, 200}   % teal cyan
\definecolor{refblue2}{RGB}{0, 190, 230}   % bright aqua
\usepackage[unicode]{hyperref}
\hypersetup{
  colorlinks=true,
  linkcolor=refblue1,
  citecolor=red,
  urlcolor=refblue2,
  pdfauthor={}
  pdftitle={}
}
\newcommand\inp[2]{\langle #1 \,|\, #2 \rangle}
\def \msun      {\rm{M}_\odot}
\def \mchirp    {\mathcal{M}}

\def \fmin      {f_{\rm min}}
\def \fmax      {f_{\rm max}}
\def \Mpc       {\mathrm{Mpc}}
\def \dL        {d_{\rm L}}

\microtypesetup{protrusion=true,expansion=true}

\usepackage[capitalise,noabbrev]{cleveref}

\crefname{figure}{Fig.}{Figs.}
\Crefname{figure}{Fig.}{Figs.}
\crefformat{figure}{Fig.\thinspace(#2#1#3)}
\Crefformat{figure}{Fig.\thinspace(#2#1#3)}
\crefname{equation}{Eq.}{Eqs.}
\Crefname{equation}{Eq.}{Eqs.}
\crefformat{equation}{Eq.\thinspace(#2#1#3)}
\Crefformat{equation}{Eq.\thinspace(#2#1#3)}
\crefname{section}{Sec.}{Secs.}
\Crefname{section}{Sec.}{Secs.}
\crefformat{section}{Sec.\thinspace#2#1#3}
\Crefformat{section}{Sec.\thinspace#2#1#3}
\usepackage{xcolor}

\usepackage{inconsolata}
\newcommand{\IMRPhenomD}{\texttt{IMRPhenomD}}

\usepackage{orcidlink}
\usepackage[normalem]{ulem} % \uline, \sout
\def \IITGn {Department of Physics, Indian Institute of Technology Gandhinagar, Gujarat 382055, India.\vspace*{3pt}}
\def \IUCAA {Inter-University Centre for Astronomy and Astrophysics (IUCAA), Pune 411007, India}

\usepackage{etoolbox}
\makeatletter
\appto{\frontmatter@authorformat}{\fontsize{9.5}{11.5}\selectfont}
\makeatother
\begin{document}
\title{Probing dark sector with decihertz gravitational wave observations of compact binary inspirals in IndIGO-D}
\author{\sc{Divya Tahelyani}\orcidlink{0000-0001-6146-7550}}
\email{divya.tahelyani@iitgn.ac.in}
\affiliation{\IITGn}
\author{\sc{Abhishek~Sharma}\orcidlink{0009-0007-2194-8633}}
\email{abhishek.sharma@iucaa.in}
\affiliation{\IUCAA}
\author{\sc{Anand~S.~Sengupta}\orcidlink{0000-0002-3212-0475}}
\email{asengupta@iitgn.ac.in}
\affiliation{\IITGn}
\author{\sc{Arpan~Bhattacharyya}\orcidlink{0000-0002-7933-6441}}
\email{abhattacharyya@iitgn.ac.in}
\affiliation{\IITGn}
\vspace*{5pt}
%\email{}
%
%
\begin{abstract}
We investigate the potential of decihertz gravitational-wave (GW) observations of compact-binary inspirals to probe atomic dark matter (aDM) and its interactions through an ultralight massive dark photon. We consider a scenario in which neutron stars accumulate DM charged under a dark Abelian massive vector mediator, leading to an additional long-range Yukawa interaction between compact objects and associated dipolar radiation. In this framework, the GW waveform acquires characteristic corrections parameterized by the Yukawa coupling strength~$\alpha$, the dipole-radiation parameter~$\gamma$, and a parameter related to the mediator mass~$\mu$.
% with dark dipole radiation entering at $-1$ post-Newtonian (PN) order, while finite-mediator-mass effects also introduce lower-order corrections.
These corrections include a leading $-1$ post-Newtonian dipole contribution and finite-mediator-mass effects.
% We first connect these aDM waveform corrections to existing parameterized tests of general relativity. Using simulated binary neutron star (BNS) events analogous to GW170817 and GW190425, and neutron star–black hole (NSBH) events analogous to GW200105, GW200115 and GW230529, we perform Bayesian parameter-estimation analyses incorporating the modified inspiral waveform in a representative decihertz detector, named IndIGO-D. 
We relate the aDM corrections to existing parameterized tests of general relativity and perform Bayesian parameter-estimation analyses for simulated binary neutron star (BNS) and neutron star--black hole (NSBH) systems incorporating the modified binary inspiral in a proposed space-based decihertz detector, named IndIGO-D.
For a GW170817-like BNS with nonzero aDM couplings chosen near current observational bounds, we find
%with nonzero aDM couplings adopted from the dedicated aDM analysis of Ref.~\cite{Owen:2025odr} as reference values, we find
% that a detectable dark-sector signal can enable the dark-photon rest energy to be inferred with an uncertainty of $\mathcal{O}(10^{-20})\,\mathrm{eV}$ at 90\% credibility with IndIGO-D.
that the dark-photon rest energy can be inferred with a $90\%$ credible uncertainty of $\mathcal{O}(10^{-20})\,\mathrm{eV}$, conditional on a detectable dark-sector contribution.
Analyses with weaker dark-sector interactions $(\alpha \lesssim \mathcal{O}(10^{-8}))$ show that the mediator mass remains unconstrained when both the Yukawa coupling and dipole-radiation contribution are negligible, but can be measured even for a negligible Yukawa coupling if the dipole-radiation contribution is sufficiently strong.
Our results demonstrate that future decihertz GW observatories can serve as powerful laboratories for testing dark-sector physics and probing dark matter interactions in compact binaries through precision gravitational-wave measurements.
\end{abstract}
\maketitle
%%================================%%
\section{Introduction}
\label{sec:introduction}
%%================================%%
%
The existence of dark matter (DM) is firmly established through a wide range of astrophysical and cosmological observations, including galactic rotation curves~\cite{Rubin:1970zza}, gravitational lensing~\cite{Clowe:2006eq}, the dynamics of galaxy clusters~\cite{Zwicky:1933gu}, measurements of the cosmic microwave background~\cite{Planck:2018vyg}, and the formation of large-scale structure~\cite{Planck:2018vyg}. Despite overwhelming evidence for its gravitational influence, the microscopic nature of DM remains one of the most fundamental unanswered questions in modern physics~\cite{Bertone:2004pz}.  In recent years, compact binary systems involving neutron stars (NSs) and black holes (BHs) have emerged as promising laboratories to probe beyond-standard-model physics, including possible signatures of DM, through gravitational-wave (GW) observations~\cite{Berti:2015itd,Yunes:2013dva}.

A particularly promising connection between GW observations and DM arises from the possibility that NSs accumulate gravitationally bound DM during their evolution in galactic environments~\cite{Goldman:1989nd, Kouvaris:2007ay, deLavallaz:2010wp, Kouvaris:2010vv, Kouvaris:2010jy, McDermott:2011jp, Guver:2012ba, Bell:2013xk, Bramante:2013hn, Bramante:2013nma}. As they traverse DM halos, NSs can capture DM particles through scattering and subsequent gravitational binding, potentially leading to a non-negligible DM component in their interiors. 
% A well-motivated scenario in this context is atomic dark matter (aDM), in which dark matter particles interact via a dark Abelian massive vector field.
A well-motivated scenario in this context is atomic dark matter (aDM), in which an asymmetric dark matter particle is charged under a hidden Abelian gauge symmetry, with interactions mediated by a massive dark vector boson (dark photon)~\cite{Kaplan:2009de,Cyr-Racine:2012tfp}. 
In this case, the accumulation of DM inside NSs can result in an effective dark charge associated with each compact object~\cite{Kopp:2018jom,Croon:2017zcu,Alexander:2018qzg}. Consequently, binary systems containing NSs can experience additional interactions mediated by the dark photon, leading to observable modifications in their dynamics and the emitted GW signal~\cite{Owen:2025odr}. A recent work has emphasized that the repulsive dark-vector interaction can itself limit the amount of net dark charge accumulated by a neutron
star, so that observable dark-electromagnetic effects depend sensitively
on the underlying particle-physics model and astrophysical environment~\cite{Harris:2025dmg}.

% A well-motivated scenario in this context is atomic dark matter (aDM), in which an asymmetric dark matter particle is charged under a hidden Abelian gauge symmetry, with interactions mediated by a massive dark vector boson (dark photon)~\cite{Kaplan:2009de,Cyr-Racine:2012tfp}. 

The resulting dark-sector interaction modifies both the conservative and dissipative sectors of the binary evolution. In the conservative sector, the additional long-range Yukawa interaction alters the total binding energy and orbital dynamics and results in the leading order contribution at $-2$ post-Newtonian (PN) order in GW phase and amplitude.
In the dissipative sector, the dark vector field opens additional radiation channels. In particular, the contribution due to electric dipole radiation arises at $-1$PN order,
% and can dominate the dark-sector dissipative correction at low frequencies,
while electric quadrupole radiation enters at Newtonian order. 
% Finite-mediator-mass effects introduce further negative-PN corrections.
%
These effects collectively introduce characteristic corrections to the GW amplitude and phase, which can be parameterized by three dimensionless quantities: the Yukawa coupling strength $\alpha$, the dipole radiation parameter $\gamma$, and the dark-photon mass parameter $\mu$~\cite{Owen:2025odr}.

An important feature of these dark-sector effects is the presence of negative-PN corrections to the waveform. Their impact, therefore, becomes increasingly important during the early inspiral, making the decihertz band particularly favorable for probing weak dark-sector interactions. BNS and NSBH systems can remain observable in this band for weeks to months, accumulating a large number of GW cycles over which small phase modifications can build up coherently~\cite{Isoyama:2018rjb, Sharma:2026tff}. Moreover, the characteristic frequencies at which dark quadrupole and dipole radiation become kinematically accessible can lie directly within the decihertz observing band. Since these activation frequencies depend on the mediator mass, their presence in the observed signal provides information about $\mu$ that is largely inaccessible to current ground-based detectors~\cite{Owen:2025odr}. Several GW observatories have been proposed to explore the decihertz frequency bands, including DECIGO~\cite{Kawamura:2006up,Sato:2017dkf,Kawamura:2020pcg}, TianGO~\cite{Kuns:2019upi}, the Lunar Gravitational-Wave Antenna (LGWA)~\cite{LGWA:2020mma, Ajith:2024mie}, the Laser Interferometer Lunar Antenna (LILA)~\cite{Jani:2025uaz,Creighton:2025kth} and IndIGO-D~\cite{Sharma:2026tff}.

In this work, we investigate the prospects for constraining dark-sector effects in compact-binary inspirals with IndIGO-D~\cite{Sharma:2026tff}, a proposed Indian decihertz GW detector concept designed to operate in the $\sim0.1$--$10\,\mathrm{Hz}$ band between millihertz space-based observatories and ground-based interferometers. We incorporate the aDM corrections derived in~\cite{Owen:2025odr} into the inspiral waveform and perform a full Bayesian parameter-estimation analysis to quantify the measurability of the dark-sector parameters $\alpha$, $\gamma$, and $\mu$, which collectively characterize the modifications induced by the dark vector interaction.
To assess the detector's capability across different source classes, we analyse simulated BNS systems analogous to GW170817 and GW190425~\cite{GW190425Discovery}, as well as neutron star--black hole (NSBH) systems analogous to GW200105, GW200115~\cite{LVC:NSBHdiscovery}, and GW230529~\cite{LIGOScientific:2024elc}. Complementary studies have also explored the potential of IndIGO-D for direct detection of ultralight DM, including scalar, vector, tensor, and axion candidates~\cite{Hoque:2026qze}.

Previous studies have investigated the signatures of such dark-vector interactions in compact-binary inspirals. Alexander \textit{et al.}~\cite{Alexander:2018qzg} developed an inspiral-only \texttt{TaylorF2} waveform including leading-PN modifications to the binary energy and radiated power from the dark Yukawa interaction and associated dipole radiation. Using Fisher-matrix analysis, they forecast that next-generation ground-based detectors could constrain the effective Yukawa coupling and dipole-radiation parameters to approximately $\alpha\lesssim\mathcal{O}(10^{-5})$ and $\gamma\lesssim\mathcal{O}(10^{-7})$ at \(1\sigma\) credibility, respectively. More recently, Owen \textit{et al.}~\cite{Owen:2025odr} extended this framework to higher PN order, incorporated the aDM corrections into the \texttt{IMRPhenomD\_NRTidalv2} waveform model~\cite{Dietrich:2019kaq}, and performed Bayesian parameter-estimation analyses using current ground-based GW observations. Their strongest constraints were obtained from GW170817, with $\alpha\lesssim\mathcal{O}(10^{-2})$ and $\gamma\lesssim\mathcal{O}(10^{-4})$ at $90\%$ credibility. However, the dark-photon mass parameter $\mu$ remained effectively unconstrained. 

In addition to dedicated dark-sector analyses, existing parameterized tests of general relativity performed by the LIGO--Virgo--KAGRA (LVK) Collaboration~\cite{LVC:GWTC3TGR,LVC:GWTC4TGR} provide an indirect means of constraining aDM interactions. We establish a correspondence between the aDM-induced phase corrections and the parameterized post-Newtonian (PN) deviations, allowing us to translate published bounds from GW170817~\cite{LVC:GW170817discovery,LVC:GW170817TGR} and GW230529~\cite{LIGOScientific:2024elc,LVC:GW230529TGR} into approximate constraints on $\alpha$ and $\gamma$. We also identify a limitation of these tests: the $-2$PN aDM contribution, proportional to $\alpha\mu^2$, is not included among the currently tested deformation parameters. The enhanced sensitivity to such negative-PN corrections at low frequencies provides further motivation for decihertz observations. We discuss these constraints in Sec.~\ref{subsec:adm_past_gw}.

Our analysis demonstrates that the decihertz band provides a substantial gain in sensitivity to aDM-induced modifications of compact-binary inspirals.  For a GW170817-like BNS system, the parameters $\alpha$, $\gamma$, and $\mu$ are recovered with $90\%$ credible uncertainties of approximately $3.0\times10^{-5}$, $1.4\times10^{-7}$, and $1.5\times10^{-9}$, respectively. The low-frequency coverage can place the dark quadrupole and dipole radiation thresholds within the observing band, providing direct information about the mediator mass. 
% A particularly important result of this study is the improved inference of the dark photon mass parameter~$\mu$. 
For the benchmark aDM signals considered here, we find that both BNS and NSBH
observations with IndIGO-D can provide informative constraints on $\mu$.
% We show that observations of both BNS and NSBH systems with IndIGO-D can place meaningful constraints on $\mu$
% , thereby opening a new observational window into the properties of dark-sector interactions.
%
Conditional on a detectable aDM contribution, we infer a dark-photon rest energy of $m_Ac^2\simeq2.42\times10^{-15}\,\mathrm{eV}$ with an absolute 90\% credible uncertainty of order $10^{-20}\,\mathrm{eV}$ for the GW170817-like benchmark system.
% For the GW170817-like benchmark system, we use this analysis as a demonstration of the achievable mediator-mass precision when the aDM parameters are themselves measurably imprinted on the waveform. Conditional on such a detectable aDM contribution, we infer a dark-photon rest energy $m_A c^2 \simeq 2.42\times10^{-15}\,\mathrm{eV}$ with an absolute $90\%$ credible uncertainty of order $10^{-20}\,\mathrm{eV}$. These results highlight the potential of future decihertz GW observatories as precision probes of aDM and other beyond-standard-model scenarios in compact binary systems.

The benchmark dark-sector interactions considered in this work are motivated by existing observational upper limits as obtained in Ref.~\cite{Owen:2025odr}. However, studies by Kopp et al.~\cite{Kopp:2018jom} and Harris and Kahn~\cite{Harris:2025dmg} have highlighted the astrophysical difficulties associated with generating and retaining sufficiently large net dark charges in NSs, suggesting that the resulting interaction strengths may be substantially smaller. Motivated by these considerations, we extend our parameter-estimation analysis to weaker dark-sector couplings for a GW170817-like BNS system. We find that for a Yukawa coupling of ${\alpha=10^{-8}}$, the mediator-mass parameter $\mu$ remains unconstrained when the dipole-radiation contribution vanishes, whereas both $\alpha$ and $\mu$ can be measured for ${\alpha\sim10^{-5}}$. We further demonstrate that a nonzero dipole-radiation contribution can allow $\mu$ to be measured even when the conservative Yukawa coupling is too weak to be resolved. These analyses highlight the complementary roles of conservative and dissipative dark-sector effects in determining the measurability of the mediator mass.

The remainder of this paper is organized as follows. In Sec.~\ref{sec:Atomic_dark_matter_model}, we summarize the aDM model and the resulting modifications to the conservative and dissipative binary dynamics, construct the frequency-domain waveform, and describe the phase-matching procedure required when the dark quadrupole and dipole activation frequencies lie within the detector band. In Sec.~\ref{subsec:adm_past_gw}, we relate the aDM phase corrections to the parameterized tests of general relativity and translate existing LVK bounds into constraints on the dark-sector parameters. In Sec.~\ref{subsec:PS_restrictions}, we describe the range of dark-sector parameters considered for decihertz observations and Sec.~\ref{sec:methodology} describes the details of Bayesian analysis setup for all the systems. Sec.~\ref{subsec:intrinsic_ps_match} provides the match-based exploration in the intrinsic parameter space used to construct suitable prior ranges. The parameter-estimation results for the representative BNS and NSBH systems, including the inferred dark-photon mass and sensitivity to weaker dark-sector interactions
% , and the timing properties of the long-duration signals,
are presented in Sec.~\ref{sec:results}. Finally, Sec.~\ref{sec:conclusion} summarizes our main findings and discusses their implications for probing dark-sector physics with future decihertz GW observatories.
%

%%================================================================%%
\section{Atomic dark matter model and the modified binary inspiral}
\label{sec:Atomic_dark_matter_model}
%%================================================================%%
%
In this section, we summarize the theoretical framework developed by Owen \textit{et al.}~\cite{Owen:2025odr}, which rigorously computes these modifications upto 1PN order. We give a brief overview of the aDM model and outline the steps to obtain corrections to the conservative and dissipative sectors of binary evolution, and finally, the modified frequency domain waveform. %We work in units with $c=\hbar=1$ unless stated otherwise.

%----------------------------%
\subsection{The aDM model}
%----------------------------%
%
%In this framework, the dark sector is composed of a Dirac fermion $\chi$ with mass $m_\chi$, which possesses a unit charge under a dark $U(1)_D$ gauge symmetry with a coupling constant $g_D$.
In the effective description adopted here, following
Ref.~\cite{Owen:2025odr}, the dark sector is represented by a Dirac
fermion $\chi$ with mass $m_\chi$ carrying charge under a dark
$U(1)_D$ gauge symmetry with coupling constant $g_D$.
This captures the charged dark-sector degrees of freedom relevant for
the compact-binary dynamics; the detailed microscopic realization of
the atomic dark-matter sector is not required for the waveform model.
The gauge boson $A^\mu$, regarded as the dark photon, is allowed to have a nonzero mass $m_A$. Thus, the dark sector Lagrangian density is given by,
\begin{align}
\mathcal{L}_{\text{DS}} = & -\frac{1}{4} F_{\mu\nu} F^{\mu\nu} - \frac{1}{2} \left(\frac{c}{\hbar} m_A\right)^2 A^\mu A_\mu \nonumber \\
& + \bar{\chi} \left( i\gamma^\mu D_\mu - \frac{c}{\hbar} m_\chi \right) \chi \, , \label{eq:lagrangian_DS}
\end{align}
where, $F_{\mu\nu} = \nabla_\mu A_\nu - \nabla_\nu A_\mu$ is the field strength tensor of the vector field and $D_\mu = \nabla_\mu + ig_D A_\mu$ is the gauge-covariant derivative. Incorporating the aDM Lagrangian density, the total Lagrangian density becomes
\begin{equation}
\mathcal{L} = \mathcal{L}_{\text{EH}} + \mathcal{L}_{\text{SM}} + \mathcal{L}_{\text{DS}} \, , \label{eq:total_lagrangian}
\end{equation}
where, $\mathcal{L}_{\text{EH}} = g^{\mu\nu} R_{\mu\nu} / (16\pi G)$ is the Einstein-Hilbert Lagrangian density of general relativity and $\mathcal{L}_{\text{SM}}$ describes the standard model matter distribution of our system. In $\mathcal{L}_{\text{EH}}$, $R_{\mu\nu}$ is the Ricci tensor associated with the metric tensor $g_{\mu\nu}$ and $G$ is the Gravitational constant. After varying the action associated with the Lagrangian density $\mathcal{L}$, one obtains the Proca equation
\begin{equation}
\nabla_\beta ( \nabla^\alpha A^\beta - \nabla^\beta A^\alpha ) + \frac{1}{\lambda^2} A^\alpha = \frac{4\pi}{c} J^\alpha \, , \label{eq:Proca_equation}
\end{equation}
and the Einstein equations
\begin{equation}
\left( R_{\mu\nu} - \frac{1}{2} \, g^{\alpha\beta} \, R_{\alpha\beta} \, g_{\mu\nu} \right) = \frac{8\pi G}{c^4} T_{\mu\nu} \, , \label{eq:Einstein_equation}
\end{equation}
where, $J^\mu = (c / 4\pi) g_D \bar{\chi} \gamma^\mu \chi$ is the dark four-current density and $\lambda = \hbar / (c m_A)$ is the reduced Compton wavelength of the dark vector field $A^\mu$, and $T_{\mu\nu} = T_{\mu\nu}^{\text{matter}} + T_{\mu\nu}^\text{A}$ is the total stress-energy tensor. The contribution of massive vector field $T_{\mu\nu}^\text{A}$ to the stress-energy tensor must be small and can be discarded, thus the massive vector field can be considered as a test field on the standard PN binary background.
The solution to the Einstein equations and the Proca field equation are obtained using the PN formalism. The modified equation of motion of the binary are then obtained by first calculating the acceleration of one of the bodies due to the Proca field and the spacetime curvature and later utilizing the problem's symmetries to calculate the acceleration of the second body. 

Following Ref.~\cite{Owen:2025odr}, the modifications to the binary dynamics
can be expressed in terms of three dimensionless dark-sector parameters.
Denoting the effective dark charges of the two compact objects by $q_1$ and
$q_2$, these parameters are
\begin{equation}
\alpha
=
\frac{\hbar c}{G}
\frac{q_1}{m_1}
\frac{q_2}{m_2},
\;
\gamma
=
\frac{\hbar c}{G}
\left(
\frac{q_1}{m_1}-\frac{q_2}{m_2}
\right)^2,
\;
\mu
=
\frac{Gmm_A}{\hbar c}.
\label{eq:adm_phenomenological_parameters}
\end{equation}
where $m = m_1+m_2$. The parameter $\alpha$ characterizes the strength of
the dark Yukawa interaction relative to gravity, while $\gamma$ quantifies
the charge-to-mass asymmetry between the two bodies and controls the
strength of the leading dipole-radiation contribution. The dark-photon mass parameter, $\mu$ sets the finite interaction range and the frequencies at which the massive dark-radiation channels become kinematically accessible.

However, to make the derivation of the two-body dynamics tractable, the
analysis in \cite{Owen:2025odr} is restricted to the \textit{long-range
limit}. This regime assumes that the orbital separation of the binary
$r$ is much smaller than the mediator's Compton wavelength,
${r \ll \lambda}$ which corresponds to ${\mu/x \ll 1}$ in terms of the waveform variables.
% In terms of the waveform variables, this corresponds to $\mu/x \ll 1$, and the expressions used below are expanded through $\mathcal{O}(\mu^2/x^2)$.
%
%However, to make the derivation of the two-body dynamics tractable, the analysis in \cite{Owen:2025odr} is restricted to the \textit{long-range limit}. This regime assumes that the orbital separation of the binary $r$ is much smaller than the mediator's Compton wavelength ($r \ll \lambda$).

%-------------------------------------%
\subsection{Modified binary inspiral}
\label{subsec:modified_binary_inspiral}
%-------------------------------------%
%
Computing the inspiral waveform requires evaluating the binary's total energy $E$ and radiated power $\mathcal{P}$. Each of these quantities consists of a purely gravitational contribution and an additional correction arising from DM interactions. The DM corrections at the leading PN order are computed in~\cite{Alexander:2018qzg} and next-to-leading-order corrections are computed in~\cite{Owen:2025odr}.
Both are expanded in the PN parameter
\begin{equation}
x \equiv \left(\frac{Gm\omega}{c^3}\right)^{2/3} = \left(\frac{G\pi m f}{c^3}\right)^{2/3}\, ,
\label{eq:x_def}
\end{equation}
where $\omega$ is the orbital angular frequency, and $f=\omega/\pi$ is the GW frequency (twice the orbital frequency). For a Newtonian circular orbit of separation $r$, Kepler's third law $\omega^2 r^3 = Gm$ together with the orbital speed $v=\omega r$ gives $v^2 = Gm/r = (Gm\omega)^{2/3}$, so that $x=v^2/c^2$ at leading order: $x$ is a book-keeping parameter for the orbital speed in units of $c$, and the PN expansion below is an expansion in powers of $(v/c)^2$.
The total conserved binding energy $E(x)$ and radiated power $\mathcal{P}(x)$ determine the modified inspiral through the energy-balance relation. Their explicit closed-form expressions are given in Appendix~\ref{app:coefficients}, Eqs.~\eqref{eq:Ex_closed} and \eqref{eq:Px_closed}, and we summarize here only the structure relevant for the waveform construction.

The conservative dynamics are modified primarily by the effective Yukawa coupling strength $\alpha$, 
% defined as the ratio of the dark electric force to the standard gravitational force, 
together with finite-range corrections controlled by the dark-photon mass parameter $\mu$. The binding energy $E(x)$ contains no Heaviside factors, since the conserved energy does not depend on whether a radiation channel is kinematically open. By contrast, the Proca field opens an electric dipole radiation channel entering at $-1$PN order and an electric quadrupole channel entering at Newtonian order, while also modifying the gravitational flux through 1PN order. Consequently, $\mathcal{P}(x)$ contains Heaviside functions marking the kinematic onset of the dipole and quadrupole dark-photon radiation channels. This distinction between the conservative and dissipative sectors will be important below when discussing the amplitude and phase across the activation thresholds.

%----------------------------------%
\subsection{Gravitational waveform}
\label{subsec:gravitational_waveform}
%----------------------------------%
%
Using the modified binding energy $E(x)$ (Eq.~\eqref{eq:Ex_closed}) and the total energy flux $\mathcal{P}(x)$ (Eq.~\eqref{eq:Px_closed}), the frequency-domain GW strain $\tilde{h}(f)$ can be computed under the stationary phase approximation (SPA)~\cite{Cutler:1994ys, Droz:1999qx}:
\begin{equation}
\tilde{h}(f) = \tilde{A}(f)e^{i\tilde{\varphi}(f)} \, , \label{eq:h(f)}
\end{equation}
We denote this total Fourier-domain phase $\tilde\varphi(f)$ throughout the present subsection. Section~\ref{subsec:adm_past_gw} instead writes $\Psi(f)$ for the accumulated inspiral phase alone, the PN series inside $\tilde\varphi(f)$ with the reference time/phase offset $2\pi f t_c-2\varphi_c-\pi/4$ set aside. The correspondence between the two is made explicit there.
The Fourier transform $\tilde h(f)=\int_{-\infty}^{\infty} A(t)e^{2\pi i f t}e^{-2i\varphi(t)}\,dt$ is dominated by the stationary point $t_0$, defined by $\dot\varphi(t_0)=\pi f\equiv\omega(t_0)$. Expanding about this point gives the general SPA relations, Eq.~(70) of Ref.~\cite{Owen:2025odr}:
\begin{subequations}
\label{eq:SPA_general}
\begin{align}
\tilde{A}(f) &= A(t_0)\left(\frac{\pi}{\dot\omega(t_0)}\right)^{1/2}\, , \\
\tilde{\varphi}(f) &= 2\pi f\,t_0 - 2\varphi(t_0) - \frac{\pi}{4}\, .
\end{align}
\end{subequations}
Equation~\eqref{eq:SPA_general} is completely general: every aDM correction enters only through $t_0(f)$, $\varphi(t_0)$, and $\dot\omega(t_0)$, which follow from the energy-balance equation ${\mathcal{P}=-dE/dt}$ together with $\omega=(c^3/Gm)x^{3/2}$,
\begin{subequations}
\label{eq:t0_phi_omegadot}
\begin{align}
t_0(x) &= t_c - \int_{x_c}^{x} \! dx'\,\frac{E'(x')}{\mathcal{P}(x')}\, , \\
\varphi(t_0) &= \varphi_c - \frac{c^3}{Gm}\int_{x_c}^{x} \! dx'\, x'^{3/2}\frac{E'(x')}{\mathcal{P}(x')}\, , \\
\dot\omega(t_0) &= -\frac{3}{2}\frac{c^3}{Gm}x^{1/2}\,\frac{\mathcal{P}(x)}{E'(x)}\, ,
\end{align}
\end{subequations}
evaluated at $x=x(t_0)$ (Eq.~\eqref{eq:x_def}), with $\eta=m_1m_2/(m_1+m_2)^2$ the symmetric mass ratio, $d_L$ the luminosity distance, $x_c\equiv x(t_c)$ the value of $x$ at coalescence, and $t_c,\varphi_c$ the coalescence time and phase. Substituting Eqs.~\eqref{eq:Ex_closed}--\eqref{eq:Px_closed} into Eq.~\eqref{eq:t0_phi_omegadot}, expanding to linear order in $\alpha,\gamma$ and to $\mathcal{O}(\mu^2/x^2)$, and collecting the result as a single series in $x$ gives
\begin{subequations}
\label{eq:h_series}
\begin{align}
\tilde{A}(f) &= -\frac{G^2}{c^5} \sqrt{\frac{5\pi}{24}} \frac{m^2\sqrt{\eta}}{d_L} \frac{1}{x^{7/4}} \sum_{n=-2}^{1} \tilde{A}_{(n)} x^n \, , \\
\tilde{\varphi}(f) &= 2\pi f t_c - 2\varphi_c - \frac{\pi}{4} + \frac{3}{128\eta} \frac{1}{x^{5/2}} \sum_{n=-2}^{1} \tilde{\varphi}_{(n)} x^n. \label{eq:h_series_phase}
\end{align}
\end{subequations}
The various coefficients in the above expansions are lengthy closed-form expressions in $\alpha,\gamma,\mu,\eta,\Delta$. We collect them in Appendix~\ref{app:coefficients} (Eq.~\eqref{eq:adm_amp_phase}) and record here only their essential physical content.
An important feature of Eq.~\eqref{eq:adm_amp_phase} is the appearance of the Heaviside functions, which determine the frequencies at which the additional radiation channels associated with the Proca field become kinematically accessible. The electric dipole and quadrupole radiation modes are activated when
\begin{subequations}
\label{eq:threshold_conditions}
\begin{align}
\frac{x^{3/2}}{\mu}&>1\, , \\
\frac{x^{3/2}}{\mu}&>\frac{1}{2}\, ,
\end{align}
\end{subequations}
respectively. Using the PN parameter $x$ of Eq.~\eqref{eq:x_def}, the corresponding activation frequencies can be written as
\begin{subequations}
\label{eq:activation_freqs}
\begin{align}
f_{\rm dipole}&=\frac{\mu c^3}{\pi G m}\, , \\
f_{\rm quadrupole}&=\frac{\mu c^3}{2\pi G m}\, ,
\end{align}
\end{subequations}
such that \(f_{\rm dipole}=2f_{\rm quadrupole}\). 
%
%Since both frequencies scale linearly with the dark-photon mass parameter \(\mu\), the locations of these spectral features provide a direct observational probe of the mediator mass. If these thresholds lie within the sensitive band of a GW detector, the onset of the corresponding radiation channels imprints characteristic signatures in the waveform amplitude and phase evolution.

Since both frequencies scale linearly with the dark-photon mass
parameter $\mu$, the locations of these spectral features provide a
direct observational probe of the mediator mass. In the PN-expanded
waveform adopted here, following Ref.~\cite{Owen:2025odr}, the
kinematic opening of the two radiation channels is represented by
Heaviside functions and therefore appear as a sharp activation at
$f_{\rm quadrupole}$ and $f_{\rm dipole}$. If these activation
frequencies lie within the sensitive band of a GW detector, their
locations imprint characteristic signatures in the waveform amplitude
and phase evolution. The mediator-mass constraints obtained below
should therefore be understood within this waveform model; a more
complete treatment of the detailed near-threshold behavior may modify
the information carried by the immediate vicinity of the activation
frequencies.

\subsubsection{Matching the phase across a threshold}
\label{subsubsec:matching}
Equation~\eqref{eq:h_series_phase} is obtained by holding ${\Theta(x^{3/2}/\mu-1)}$ and ${\Theta(x^{3/2}/\mu-\tfrac12)}$ - the threshold conditions of Eq.~\eqref{eq:threshold_conditions} - fixed while carrying out the integrals of Eq.~\eqref{eq:t0_phi_omegadot}. 
%
%This is exact provided a detector's band lies entirely on one side of both thresholds, 
Within this piecewise waveform model, this treatment is
self-consistent provided a detector's band lies entirely on one side
of both thresholds, 
which holds for every ground-based analysis to date, including the original analysis of Ref.~\cite{Owen:2025odr}: there, $\mu<10^{-4}$ places $f_{\rm dipole}$ and $f_{\rm quadrupole}$ below the seismic wall. For IndIGO-D this condition fails. As shown in Sec.~\ref{sec:Probing_aDM_in_neutron_stars_in_dHz_band}, both activation frequencies lie inside the observing band for the fiducial $\mu$ used in this work, so the integration range in Eq.~\eqref{eq:t0_phi_omegadot} crosses a threshold. Therefore, $\mathcal{P}(x)$ (Eq.~\eqref{eq:Px_closed}), and the integrand $g(x)\equiv E'(x)/\mathcal{P}(x)$, changes discontinuously. This directly affects \(t_0(x)\), \(\varphi(t_0)\), and therefore the accumulated Fourier-domain phase, requiring a consistent treatment of the waveform phase across the corresponding thresholds.
% This has direct implications for $t_0(x)$ and $\varphi(t_0)$, and hence for the total accumulated phase. Thus, we require a consistent treatment of the waveform phase across the corresponding thresholds.

Let $G_k$ denote an antiderivative of $g$ on region $k$, so that
\begin{equation}
\int_{x_k}^{x}\!dx'\,g(x')\Big|_{\text{region }k} = G_k(x)-G_k(x_k)
\end{equation}
for any reference point $x_k$ inside region $k$. Define $H_k$ in the same way, through $H_k'(x)\equiv x^{3/2}g(x)|_{\text{region }k}$, for the second integral of Eq.~\eqref{eq:t0_phi_omegadot}. On region $k$, Eq.~\eqref{eq:t0_phi_omegadot} then reads
\begin{subequations}
\begin{align}
t_0(x)\big|_k &= t_c - G_k(x) + G_k(x_k)\, , \\
\varphi(t_0)\big|_k &= \varphi_c - \frac{c^3}{Gm}\Big[H_k(x) - H_k(x_k)\Big]\, .
\end{align}
\end{subequations}

If $x$ and $x_c$ lie in the same region, the natural choice of reference point is $x_k=x_c$, for which $G_k(x_c)=H_k(x_c)=0$ by definition of a definite integral, and both expressions above reduce to Eq.~\eqref{eq:t0_phi_omegadot} itself. This is the situation in every ground-based analysis referenced above, and it is why no ambiguity arises there. 
Once $x$ has crossed into a region other than the one containing $x_c$, however, $g$ takes a different closed form on either side of the threshold, so $x_k\neq x_c$ on that region, and $G_k(x_k)$, $H_k(x_k)$ cannot be read off from Eq.~\eqref{eq:t0_phi_omegadot} directly: they are undetermined constants of integration, fixed only by matching across the threshold.

\begin{figure*}[!ht]
\centering
\includegraphics[width=0.75\textwidth]{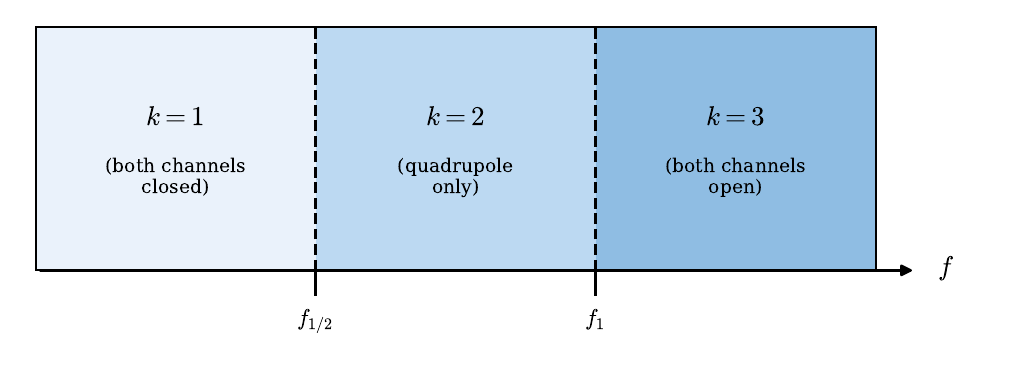}
\caption{Schematic of the three frequency regions used in the phase matching, labeled by the region index $k$ of Eq.~\eqref{eq:affine_match}. Region $k=1$ ($f<f_{1/2}$) has both the dipole and quadrupole channels kinematically closed; region $k=2$ ($f_{1/2}<f<f_1$) has only the quadrupole channel open; region $k=3$ ($f>f_1$) has both channels open. The threshold frequencies $f_{1/2}\equiv f_{\rm quadrupole}$ and $f_1\equiv f_{\rm dipole}$ (Eq.~\eqref{eq:activation_freqs}) are named after the Heaviside arguments $\theta_{1/2}^{(k)}$ and $\theta_1^{(k)}$ of Table~\ref{tab:regions}, and correspond to the thresholds $x_{\rm quad}$ and $x_{\rm dip}$ of Eq.~\eqref{eq:x_thresholds}. Region 3 contains coalescence and is anchored by $A_3=B_3=0$ in Eq.~\eqref{eq:AkBk}.}
\label{fig:zones_schematic}
\end{figure*}

Substituting $t_0(x)|_k$ and $\varphi(t_0)|_k$ into the SPA relation of Eq.~\eqref{eq:SPA_general}, the terms depending on $x_k$ collect into a single constant multiplying $2\pi f$ and a single additive constant. We call these
\begin{equation}
B_k \equiv G_k(x_k)\, , \qquad A_k \equiv -\frac{2c^3}{Gm}\,H_k(x_k)\, ,
\end{equation}
so that, on region $k$, the SPA phase becomes
\begin{equation}
\tilde\varphi(f) \to \tilde\varphi(f) + A_k + 2\pi B_k f\, ,
\label{eq:affine_match}
\end{equation}
where $\tilde\varphi(f)$ on the right is given by Eq.~\eqref{eq:h_series_phase} evaluated with the Heaviside functions fixed to their values in the region $k$.
%evaluated with $\Theta\to\theta^{(k)}$. 
Evaluating Eq.~\eqref{eq:h_series_phase} on region $k$ without this correction implicitly assumes $x_k=x_c$ on every region, which sets $A_k=B_k=0$ everywhere -- an assumption that fails once $x$ has crossed into a region other than the one containing $x_c$. 

% For the parameters used in Fig.~\ref{fig:phase_dephasing}, omitting $A_k$ and $B_k$ under these circumstances produces a spurious discontinuity in $\tilde\varphi(f)$ of order $10^4$--$10^5$\,rad at each threshold. This discontinuity is an artifact of that omission rather than a physical effect, since $t_0(x)$ and $\varphi(t_0)$ remain definite integrals of a function with, at worst, finite jump discontinuities, and a definite integral of such a function is itself continuous.

We now make this concrete for the three regions relevant to this work. As shown schematically in Fig.~\ref{fig:zones_schematic}, let region $k=1,2,3$ denote, respectively, $x<x_{\rm quad}$ (both channels closed), $x_{\rm quad}<x<x_{\rm dip}$ (quadrupole channel open only), and $x>x_{\rm dip}$ (both channels open), with thresholds
\begin{subequations}
\label{eq:x_thresholds}
\begin{align}
x_{\rm quad}&=\Big(\frac{\mu}{2}\Big)^{2/3}\, , \\
x_{\rm dip}&=\mu^{2/3}\, ,
\end{align}
\end{subequations}
obtained by solving the threshold conditions of Eq.~\eqref{eq:threshold_conditions} for $x$, consistent with $f_{\rm quadrupole}, f_{\rm dipole}$ of Eq.~\eqref{eq:activation_freqs} above. 
% On region $k$, ${\Theta(x^{3/2}/\mu-1)}$ and ${\Theta(x^{3/2}/\mu-\tfrac12)}$ take the constant values $\theta_1^{(k)}$ and $\theta_{1/2}^{(k)}$ listed in Table~\ref{tab:regions}, together with the corresponding $A_k$ and $B_k$. 
On region $k$, ${\Theta(x^{3/2}/\mu-1)}$ and ${\Theta(x^{3/2}/\mu-\tfrac12)}$ take constant values $\theta_1^{(k)}$ and $\theta_{1/2}^{(k)}$ determined by whether the dipole and quadrupole radiation channels are open or closed.
The constants $A_k$ and $B_k$ are not free parameters: they are chosen
to enforce continuity of the Fourier phase and its first frequency
derivative across each activation threshold, thereby removing the
spurious phase jumps introduced by the piecewise integration of the
sharp-threshold flux.
%
%These are not free parameters: they are fixed by requiring that $\tilde\varphi(f)$ and $d\tilde\varphi/df$ agree at each threshold, anchored by $A_3=B_3=0$ in the region containing coalescence, where $x_k=x_c$ and Eq.~\eqref{eq:h_series_phase} is already correctly normalized. 
The derivative condition is equivalent to continuity of $t_0(f)$, since $d\tilde\varphi/df=2\pi t_0(f)$ follows directly from Eq.~\eqref{eq:SPA_general} together with $\dot\varphi(t_0)=\omega(t_0)=\pi f$.

Only $\tilde\varphi_{(-1)}$ (Eq.~\eqref{eq:adm_phim1}) and $\tilde\varphi_{(0)}$ (Eq.~\eqref{eq:adm_phi0}) in Eq.~\eqref{eq:adm_amp_phase} carry $\Theta$ factors, so only these two coefficients change across a threshold, and only they enter the matching conditions above. Because $x_{\rm quad}$ and $x_{\rm dip}$ are clean powers of $\mu$, solving those conditions for $A_k$ and $B_k$ requires no root-finding: the result is closed form. Writing $C\equiv3/(128\eta)$ for the prefactor of Eq.~\eqref{eq:h_series} and $M\equiv Gm/c^3$, the matching conditions give
\begin{subequations}
\label{eq:AkBk}
\begin{align}
B_2 &= -\frac{CM}{6}\left[-\frac{5}{12}\gamma\,\mu^{-10/3} + 5\,\Xi\,\mu^{-8/3}\right] , \\
A_2 &= C\left[-\frac{25}{126}\gamma\,\mu^{-7/3} + \frac{8}{3}\,\Xi\,\mu^{-5/3}\right] , \\
B_1 &= B_2 - \frac{5\cdot2^{5/3}}{3}\,CM\,\mu^{-8/3}\,\Upsilon\, ,\\
A_1 &= A_2 + \frac{8\cdot2^{5/3}}{3}\,C\,\mu^{-5/3}\,\Upsilon\, ,
\end{align}
\end{subequations}
with $A_3=B_3=0$ and
\begin{subequations}
\begin{align}
\Xi &\equiv \alpha\frac{5\Delta}{16} - \gamma\Big(\frac{4555}{8064}+\frac{5}{9}\eta-\frac{5}{32}\Delta\Big)\, , \\
\Upsilon &\equiv -\alpha\frac{1-\Delta}{4} - \gamma\frac{(1-\Delta)^2}{16}\, .
\end{align}
\end{subequations}

\begin{table}[t]
\caption{Heaviside values and matching constants for the three frequency regions of Eq.~\eqref{eq:affine_match}, evaluated for the GW170817-like BNS injection of Fig.~\ref{fig:phase_dephasing} ($m_1=1.3998\,M_\odot$, $m_2=1.3506\,M_\odot$, $\alpha=0.035$, $\gamma=4.1\times10^{-4}$, $\mu=5\times10^{-5}$).}
\label{tab:regions}
\begin{ruledtabular}
\renewcommand{\arraystretch}{1.30}
\begin{tabular}{l c c c c}
Region $k$ & $\theta_1^{(k)}$ & $\theta_{1/2}^{(k)}$ & $A_k$ (rad) & $B_k$ (s) \\
\colrule
1 ($f<f_{\rm quad}$)          & 0 & 0 & $-1.84\times10^{5}$ & $2.50\times10^{4}$ \\
2 ($f_{\rm quad}<f<f_{\rm dip}$) & 0 & 1 & $-8.32\times10^{4}$ & $7.88\times10^{3}$ \\
3 ($f>f_{\rm dip}$)            & 1 & 1 & $0$ & $0$ \\
\end{tabular}
\end{ruledtabular}
\end{table}
%
% Table~\ref{tab:regions} lists the numerical values for the injection used in Fig.~\ref{fig:phase_dephasing}. 
Table~\ref{tab:regions} summarizes the three frequency regions, the corresponding Heaviside values $\theta_1^{(k)}$ and $\theta_{1/2}^{(k)}$, and the numerical matching constants $A_k$ and $B_k$ for the fiducial GW170817-like injection used to illustrate the aDM waveform behavior in Fig.~\ref{fig:phase_dephasing}. A more detailed discussion of the Fig.~\ref{fig:phase_dephasing} is given in Sec~\ref{subsec:PS_restrictions}.

The constants $A_k$ depend only on $(\alpha,\gamma,\mu,\eta,\Delta)$ and carry no mass scale; the total mass enters $B_k$ only through $M=Gm/c^3$. As $\mu\to0$, or as $f_{\min}$ is raised above $f_{\rm dipole}$, the thresholds $x_{\rm quad}$ and $x_{\rm dip}$ move below the observing band, the entire band lies in region 3, and $A_k,B_k\to0$ identically: Eq.~\eqref{eq:h_series_phase} is then recovered without modification, consistent with its direct use in the ground-based-detector analysis of Sec.~\ref{subsec:adm_past_gw}. 
% The dephasing shown in the lower panel of Fig.~\ref{fig:phase_dephasing} is computed from the corrected phase, Eq.~\eqref{eq:affine_match}. Using Eq.~\eqref{eq:h_series_phase} directly across $f_{\rm quad}$ and $f_{\rm dip}$ would instead introduce the two spurious jumps described above.

\subsubsection{The Fourier amplitude}
\label{subsubsec:amplitude_discontinuity}
Unlike the phase, the Fourier amplitude requires no such matching: the amplitude $\tilde A(f)=A(t_0)\sqrt{\pi/\dot\omega(t_0)}$ in Eq.~\eqref{eq:SPA_general} is a pointwise algebraic evaluation of $\dot\omega(t_0)\propto\mathcal{P}(x)/E'(x)$ at $x=x(t_0)$ (the third relation of Eq.~\eqref{eq:t0_phi_omegadot}), not an integral. There is therefore no antiderivative, and no integration constant to lose or restore. When $\mathcal{P}(x)$ (Eq.~\eqref{eq:Px_closed}) steps at $f_{\rm quad}$ or $f_{\rm dip}$ (Eq.~\eqref{eq:activation_freqs}), $\tilde A(f)$ inherits a step of the same relative size. This is expected: a radiation channel that is kinematically closed below threshold and open above it turns on discontinuously in a model that treats the onset as an instantaneous step. The aDM amplitude is therefore piecewise continuous by construction, with genuine jumps of a few percent at $f_{\rm quad}$ and $f_{\rm dip}$ for the parameters considered here. Unlike the phase discontinuity of Sec.~\ref{subsubsec:matching}, this feature cannot be removed within the present model without altering the physics it encodes.

The sharp Heaviside is itself an approximation to the true threshold behavior. As for any massive quantum, emission of a dark photon of mass $m_A$ is kinematically forbidden below $\hbar\omega=m_Ac^2$. Immediately above threshold, however, the emitted quantum is only marginally relativistic, and the radiated power should rise continuously from zero rather than jump directly to full strength. In the underlying massive-vector multipole radiation calculation of Krause, Kloor, and Fischbach~\cite{Krause:1994ar}, this appears as a phase-space factor multiplying the Heaviside, schematically ${(1-\mu^2/x^3)^{n+1/2}\,\Theta(x^{3/2}-\mu)}$. Ref.~\cite{Owen:2025odr} approximates this factor by its far-from-threshold limit of unity, i.e., by the bare $\Theta$ factors already appearing in Eq.~\eqref{eq:Px_closed}. This approximation is excellent when the thresholds sit far below the analysis band, as in current ground-based searches, but not when the band straddles them by design, as IndIGO-D does here. Recomputing $\mathcal{P}(x)$ with the exact prefactor would soften these amplitude jumps into a continuous rise over a finite bandwidth, and would render the phase matching of Sec.~\ref{subsubsec:matching} unnecessary in the first place, since $E'(x)/\mathcal{P}(x)$ would already be continuous. This refinement lies beyond the leading-order aDM waveform of Ref.~\cite{Owen:2025odr} used throughout this work; we leave it for future study. All results below therefore retain the piecewise amplitude as published, with the phase corrected as in Eq.~\eqref{eq:affine_match}.

%%============================================================%%
\section{Constraints on atomic dark matter from parameterized tests of general relativity}
\label{subsec:adm_past_gw}
%%============================================================%%
%
The aDM corrections in Eq.~\eqref{eq:adm_amp_phase} enter the inspiral phase at $-2$PN, $-1$PN, Newtonian, and 1PN orders, the latter three of which are probed by the parameterized deformation tests of the LVK Collaboration \cite{LVC:GW170817TGR,LVC:GWTC3TGR,LVC:GWTC4TGR}; no $-2$PN deformation has ever been included, a point to which we return. These tests write the frequency-domain inspiral phase in the TaylorF2 form
\begin{equation}
\Psi(f) = \frac{3}{128\eta}\, x^{-5/2} \sum_i \varphi_i^{\rm GR}\, x^{i/2}\, ,
\label{eq:TaylorF2_psi}
\end{equation}
% % where the index $i$ labels the $(i/2)$PN order. Here $\Psi(f)$ is the accumulated inspiral phase alone, not the full SPA phase $\tilde\varphi(f)$ of Eq.~\eqref{eq:h_series_phase}. It corresponds to the PN series $\sum_n\tilde\varphi_{(n)}x^n$ inside $\tilde\varphi(f)$, carrying the same overall prefactor $(3/128\eta)\,x^{-5/2}$, but omitting the reference time/phase offset $2\pi f t_c-2\varphi_c-\pi/4$ that also appears there. That offset is fit separately as an extrinsic parameter in matched-filtering searches and plays no role in the coefficient-by-coefficient deformations below.
where the index $i$ labels the $(i/2)$PN order. Here $\Psi(f)$ is the accumulated inspiral phase that corresponds to the last term in full SPA phase $\tilde\varphi(f)$ of Eq.~\eqref{eq:h_series_phase}, omitting the reference time/phase offset ${2\pi f t_c-2\varphi_c-\pi/4}$. That offset is fit separately as an extrinsic parameter in matched-filtering searches and plays no role in the coefficient-by-coefficient deformations below.

Beyond this offset, the two series differ only in how the summation index is attached to the PN order. Equation~\eqref{eq:h_series} labels its coefficients $\tilde\varphi_{(n)}$ directly by PN order $n\in\{-2,-1,0,1\}$, as in Eq.~\eqref{eq:adm_amp_phase}, while the parameterized-test literature labels its coefficients $\varphi_i^{\rm GR}$ by $i=2n$. A coefficient's numeral subscript therefore means different things in the two conventions: $\delta\hat\varphi_{-2}$ below is a deviation at $-1$PN ($n=-1$) and not $-2$PN, despite sharing the same subscript as the $-2$PN quantity $\tilde\varphi_{(-2)}$ of Eq.~\eqref{eq:adm_amp_phase}. The two conventions are distinguished typographically throughout: aDM coefficients carry a parenthesized PN-order label, $\tilde\varphi_{(n)}$, while parameterized-test coefficients carry a bare index, $\varphi_i^{\rm GR}$ and $\delta\hat\varphi_i$. Each TaylorF2 coefficient is allowed to deviate from its GR value through
\begin{equation}
\varphi_i^{\rm GR} \to (1+\delta\hat{\varphi}_i)\,\varphi_i^{\rm GR}\, ,
\label{eq:TaylorF2_deform}
\end{equation}
with the deformation treated instead as an additive constant at orders
where the GR coefficient vanishes, such as the $-1$PN dipole order
$i=-2$. Each $\delta\hat{\varphi}_i$ is then bounded by the data,
varied one at a time with all other coefficients held to their GR
values.
\begin{table*}[t]
\caption{Translation of published parameterized-test bounds into aDM
constraints through Eq.~\eqref{eq:adm_inversion}, compared with the
dedicated Bayesian fits of Ref.~\cite{Owen:2025odr} where available.
All bounds at $90\%$ credibility. The GW230529 entry assumes a black-hole primary, in which case $\alpha = 0$ exactly and the $-1$PN
inversion constrains $\gamma$ alone. The GW200115 input is inferred
from the factor $\sim 17$ improvement quoted in
Ref.~\cite{LVC:GW230529TGR}.}
\label{tab:adm_translation}
\begin{ruledtabular}
\renewcommand{\arraystretch}{1.35}
\begin{tabular}{l c c c c}
Event & Class & LVK input ($90\%$) & Implied aDM bound &
Dedicated fit~\cite{Owen:2025odr} \\
\colrule
GW170817 & BNS & $-\delta\hat{\varphi}_{-2} \leq 6.9\times10^{-6}$
 & $\gamma \lesssim 1.2\times10^{-4}$
 & $\gamma \lesssim \mathcal{O}(10^{-4})$ \\
GW170817 & BNS & $|\delta\hat{\varphi}_{2}| \approx 0.2$
 & $\alpha \lesssim 1\times10^{-1}$
 & $\alpha \lesssim \mathcal{O}(10^{-2})$ \\
GW200115 & NSBH & $|\delta\hat{\varphi}_{-2}| \sim 1.4\times10^{-3}$
 & $\gamma \lesssim 2\times10^{-2}$
 & $\gamma \lesssim \mathcal{O}(10^{-1})$ \\
GW230529 & NSBH & $|\delta\hat{\varphi}_{-2}| \lesssim 8\times10^{-5}$
 & $\gamma \lesssim 1.3\times10^{-3}$
 & (not analyzed) \\
any & --- & no $\delta\hat{\varphi}_{-4}$ exists
 & $\alpha\mu^{2}$ unconstrained & --- \\
\end{tabular}
\end{ruledtabular}
\end{table*}

The derivation of Eq.~\eqref{eq:adm_amp_phase} assumes that the reduced
Compton wavelength of the dark photon exceeds the orbital separation
throughout the observed inspiral. For a signal entering the band at
$10\,{\rm Hz}$ with ${m \gtrsim 2\,M_\odot}$, this condition translates
into the prior ${\mu < 10^{-4}}$ \cite{Owen:2025odr}. Equating the aDM coefficients $\tilde{\varphi}_{(n)}$ with their deformed GR counterparts and dropping the $\alpha\mu^{2}$ term at $-1$PN, negligible for $\mu < 10^{-4}$, we obtain
\begin{subequations}
\label{eq:adm_inversion}
\begin{align}
\delta\hat{\varphi}_{-2} &\simeq -\frac{5}{84}\,\gamma\,,
\label{eq:adm_inv_m2} \\
\delta\hat{\varphi}_{0} &\simeq \frac{5}{12}\,\alpha - 0.77\,\gamma\,,
\label{eq:adm_inv_0} \\
\delta\hat{\varphi}_{2} &\simeq 1.90\,\alpha\,.
\label{eq:adm_inv_2}
\end{align}
\end{subequations}
Equation~\eqref{eq:adm_inv_m2} carries no dependence on the masses and
holds for any binary. The coefficients in Eqs.~\eqref{eq:adm_inv_0} and
\eqref{eq:adm_inv_2} depend on the mass ratio and are quoted for equal
masses, appropriate for the BNS events that bound $\alpha$; for
GW170817 they shift by less than a percent.
Equations~\eqref{eq:adm_inv_m2} and \eqref{eq:adm_inv_2} each involve a
single aDM parameter and serve as our working relations: a $-1$PN bound
converts directly to $\gamma$ and a 1PN bound to $\alpha$. The three
relations together carry a stronger statement. According to the aDM
model, the deviation vector
$(\delta\hat{\varphi}_{-2},\,\delta\hat{\varphi}_{0},\,\delta\hat{\varphi}_{2})$
is not free to point anywhere in its three-dimensional space but must
lie in the plane spanned by the $\gamma$ direction
${(-5/84,\,-0.77,\,0)}$ and the $\alpha$ direction ${(0,\,5/12,\,1.90)}$.
Eliminating $\alpha$ and $\gamma$ expresses this as the consistency
condition
${\delta\hat{\varphi}_{0} \simeq 0.22\,\delta\hat{\varphi}_{2}
+ 12.9\,\delta\hat{\varphi}_{-2}}$ from their joint measurement, valid at the current PN order of
the aDM waveform. 

Note that LVK bounds on catalog events are derived from
a template that deviates from GR at a single PN order. If a deviation were
present, the best-fit $\delta\hat{\varphi}_i$ would therefore recover
the aDM coefficient at that order plus the projections of the other
three orders' deviations onto it, rather than the coefficient alone
\cite{LVC:GWTC3TGR, LVC:GWTC4TGR}. Since the data are consistent with GR, the
published bounds remain valid upper limits on each coefficient
separately, up to the weak cross-order correlations indicated by the
principal component analysis of Ref.~\cite{Owen:2025odr}.
Only events containing at least one neutron star carry information
about the aDM parameters as the no-hair theorems~\cite{Israel:1967wq, Israel:1967za, Carter:1971zc} forbid a net dark charge
on a black hole, so $\alpha$ and $\gamma$ vanish identically for binary
black holes, and the catalog-combined bounds of
Refs.~\cite{LVC:GWTC3TGR,LVC:GWTC4TGR}, dominated by such events, place
no constraint on the aDM model.

For GW170817, the LVK analysis reports
the one-sided bound $-\delta\hat{\varphi}_{-2} \leq 6.9\times10^{-6}$ at $90\%$ credibility \cite{LVC:GW170817TGR},\footnote{Reference~\cite{LVC:GW170817TGR} quotes this constraint as a bound on the dipole flux parameter, $B \leq 1.2\times10^{-5}$, defined through ${\mathcal{F} = \mathcal{F}_{\rm GR}\, \left(1 + B\,c^{2}/v^{2}\right)}$ and restricted to $B \geq 0$; the value above follows from the leading-order conversion $\delta\hat{\varphi}_{-2} = -4B/7$ given therein. Since $\gamma > 0$ requires $\delta\hat{\varphi}_{-2} < 0$, this is the side of the posterior that bounds $\gamma$.} which Eq.~\eqref{eq:adm_inv_m2} converts to $\gamma \lesssim 1.2\times10^{-4}$, in agreement with the $\gamma \lesssim \mathcal{O}(10^{-4})$ obtained from the Bayesian analysis of Ref.~\cite{Owen:2025odr}. A corresponding bound follows from GW230529, whose analysis reports $|\delta\hat{\varphi}_{-2}| \lesssim 8\times10^{-5}$ assuming a black-hole primary \cite{LVC:GW230529TGR}; then $\alpha = 0$ exactly and Eq.~\eqref{eq:adm_inv_m2} yields $\gamma \lesssim 1.3\times10^{-3}$. Because GW230529 postdates the event list of Ref.~\cite{Owen:2025odr}, this improves the best NSBH constraint on $\gamma$, previously $\mathcal{O}(10^{-1})$, by roughly two orders of magnitude. 

Bounds on $\alpha$ require two charged bodies and therefore rest on the binary neutron star events. For GW170817, the 1PN and Newtonian bounds are published only graphically; reading $|\delta\hat{\varphi}_{2}| \approx 0.2$ and $|\delta\hat{\varphi}_{0}| \lesssim 0.1$ from Fig.~2 of Ref.~\cite{LVC:GW170817TGR}, Eq.~\eqref{eq:adm_inv_2} gives $\alpha \lesssim 1\times10^{-1}$, while the Newtonian order, partially degenerate with the chirp mass, provides the consistent but weaker $\alpha \lesssim 2\times10^{-1}$. These estimates are weaker than the $\alpha \lesssim \mathcal{O}(10^{-2})$ of the dedicated fit in Ref.~\cite{Owen:2025odr}, which uses the correlated deviations at all four orders at once, retains the aDM corrections to the waveform amplitude that phase-only tests discard.
%, and imposes the physical priors $\alpha,\gamma \geq 0$ and $\mu < 10^{-4}$ in place of flat priors.
Translating generic deformation bounds into theory constraints in this way follows the parameterized post-Einsteinian program \cite{Yunes:2009ke, Tahura:2018zuq, Perkins:2020tra}, in which the aDM dipole
term occupies the same exponent as scalar-tensor dipoles, and it
complements model-specific searches for dark charges in NSs
\cite{Croon:2017zcu,Kopp:2018jom,Alexander:2018qzg,Dror:2019uea,Liu:2024rcm}.
Following Ref.~\cite{Harris:2025dmg}, the resulting bounds are best
read as constraints on the effective charges of the merging stars,
agnostic to the capture history that produced them.

The $-2$PN coefficient $\tilde{\varphi}_{(-2)} = (10/27)\,\alpha\mu^2$
has no counterpart in the LVK parameterized tests,
%where the only negative-PN deformation is $\delta\hat{\varphi}_{-2}$ (at $-1$PN, not
%$-2$PN, in the index convention noted after Eq.~\eqref{eq:TaylorF2_psi}), 
therefore, the $\alpha\mu^{2}$ direction is untested by construction. Nor do the
translations above constrain $\mu$ itself: with both Heaviside
functions equal to unity, $\mu$ enters Eq.~\eqref{eq:adm_amp_phase}
only through $\alpha\mu^{2}$ and $\alpha\mu$, so a single-order bound
fixes a product of $\alpha$ and $\mu$. Allowing several
$\delta\hat{\varphi}_i$ to vary at once does not alleviate the problem, since such
posteriors are known to be uninformative~\cite{LVC:GWTC3TGR} in the
LIGO band.

By contrast, Bayesian inference with the aDM waveform suffers from
neither limitation: the fit varies only $\alpha$, $\gamma$, and $\mu$,
and a given choice of these three parameters shifts all four PN orders
together, in the proportions dictated by
Eq.~\eqref{eq:adm_amp_phase}. Even so, LIGO-band observations do not constrain $\mu$:
Ref.~\cite{Owen:2025odr} finds no bound beyond the validity prior
$\mu < 10^{-4}$.
IndIGO-D aids this in two ways. First, for a fixed
$\mu$, the corrections enter as $\mu/x$ and $\mu^{2}/x^{2}$, and
$x \propto f^{2/3}$ is roughly twenty times smaller at decihertz
frequencies than in the LIGO band. The same source therefore produces
corrections larger by factors of $\sim 20$ and $\sim 500$,
respectively, with the $-2$PN phase term, growing as $x^{-2}$,
benefiting the most; light dark photons that leave marginal imprints
in the LIGO band become measurable at decihertz frequencies, subject
to the condition $\mu \ll x$ holding at the lowest
observed frequency. Second, the dipole and quadrupole radiation
channels can turn on within the band, at frequencies proportional to
$\mu$, so the turn-on frequency itself provides a direct measure of
the dark photon mass.
Table~\ref{tab:adm_translation} summarizes the translations.

%%=======================================================================================%%
\section{Probing \NoCaseChange{aDM} with compact binary inspirals involving neutron stars}
\label{sec:Probing_aDM_in_neutron_stars_in_dHz_band}
%%=======================================================================================%%
%
%-----------------------------------------------------------%
\subsection{Restrictions on the dark-sector parameter space}
\label{subsec:PS_restrictions}
%-----------------------------------------------------------%
%
The waveform model described in Sec.~\ref{sec:Atomic_dark_matter_model} is derived under the assumption that the orbital separation of the binary remains much smaller than the Compton wavelength of the dark vector mediator, $r\ll\lambda$. Expressed in terms of the dimensionless PN parameter $x$ (Eq.~\eqref{eq:x_def}) and the dark-photon mass parameter $\mu$, this condition becomes $\mu\ll x$. Since the PN parameter increases monotonically with frequency, the most stringent constraint is obtained at the lowest frequency accessible to the detector. This requirement restricts the dark-photon mass parameter to
\begin{equation}
\mu\ll2.3\times10^{-4}
\left(\frac{f}{0.1\,\mathrm{Hz}}\right)^{2/3}
\left(\frac{m}{2M_\odot}\right)^{2/3}.
\end{equation}
Throughout this work, we consider NSs with masses in the range
$1\,M_\odot<m_{1,2}<3\,M_\odot$, corresponding to total masses
$2\,M_\odot<m<6\,M_\odot$. The most restrictive condition is therefore obtained for the minimum total mass,
$m=2\,M_\odot$, at the lower edge of the IndIGO-D observing band,
$f_{\rm min}=0.1\,\mathrm{Hz}$, yielding
\begin{equation}
\mu\ll2.3\times10^{-4},
\ \text{equivalently}\ 
m_A c^2 \ll3.1\times10^{-15}\,\mathrm{eV}.
\end{equation}
% Accordingly, we restrict our analysis to this region of parameter space and we adopt the fiducial value $\mu=5\times10^{-5}$.
%Accordingly, all injection and recovery analyses presented in this work adopt the fiducial value \(\mu=5\times10^{-5}\), which satisfies the above validity condition.
Accordingly, all injection and recovery analyses
presented in this work adopt the fiducial value
$\mu = 5 \times 10^{-5}$, which lies below this
conservative validity scale. 
%For the event-specific lower cutoff frequencies used in the analyses, the corresponding values of $x$ are still larger.
For this choice of $\mu$, the electric quadrupole and dipole radiation channels become active at
$f_{\rm quad}\simeq0.59\,\mathrm{Hz}$ and
$f_{\rm dip}\simeq1.17\,\mathrm{Hz}$ (Eq.~\eqref{eq:activation_freqs}), respectively, for a GW170817-like BNS system. Both activation frequencies lie within the IndIGO-D observing band, allowing the onset of these radiation channels to imprint characteristic signatures on the GW phase evolution and thereby providing a direct probe of the dark-photon mass. 
\begin{figure}
    \centering
    \includegraphics[width=1\linewidth]{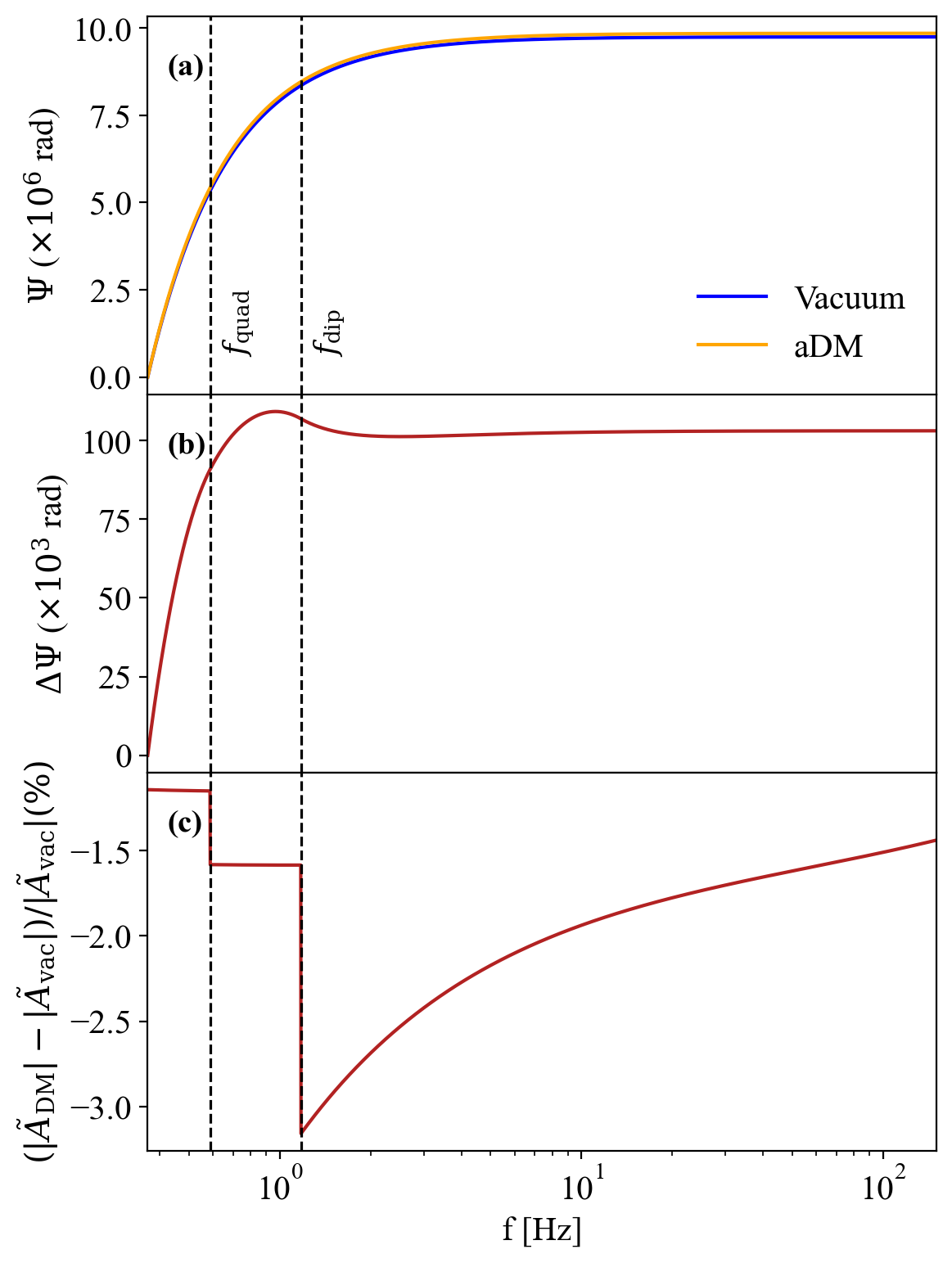}
    \caption{GW phase and amplitude for a GW170817-like binary neutron-star system ($m_1=1.3998\,M_\odot$, $m_2=1.3506\,M_\odot$, luminosity distance $40.4\,\mathrm{Mpc}$), computed from $E(x)$ and $\mathcal{P}(x)$ (Eqs.~\eqref{eq:Ex_closed}--\eqref{eq:Px_closed}) over the illustrative frequency range, $0.3648$ -- $150\,\mathrm{Hz}$. Orange curves include aDM interactions with $\alpha=0.035$, $\gamma=4.1\times10^{-4}$, and $\mu=5\times10^{-5}$; blue curves correspond to the vacuum ($\alpha=\gamma=0$) waveform. Panel (a) shows the accumulated phase, referenced to zero at $f_{\rm min}$, computed from the region-matched phase of Eq.~\eqref{eq:affine_match} (Sec.~\ref{subsubsec:matching}); the aDM curve visibly leads the vacuum curve once both channels activate. Panel (b) isolates this effect, showing the dephasing $\Delta\Psi=\Psi_{\rm DM}-\Psi_{\rm vac}$, only a percent-level fraction of the total accumulated phase in panel (a). Panel (c) shows the fractional amplitude deviation $(|\tilde A_{\rm DM}|-|\tilde A_{\rm vac}|)/|\tilde A_{\rm vac}|$: stripping the power-law envelope of $|\tilde A(f)|$ (Eq.~\eqref{eq:h_series}) exposes the aDM amplitude as manifestly \emph{piecewise continuous}, with step discontinuities at $f_{\rm quad}$ and $f_{\rm dip}$, as discussed in Sec.~\ref{subsubsec:amplitude_discontinuity}. The vertical dashed lines mark the quadrupole and dipole activation frequencies of Eq.~\eqref{eq:activation_freqs}. Discussed further in Sec.~\ref{sec:Probing_aDM_in_neutron_stars_in_dHz_band}.}
    \label{fig:phase_dephasing}
\end{figure}

Adopting the perturbative validity
criterion introduced in Ref.~\cite{Owen:2025odr}, we retain only
terms linear in $\alpha$ and $\gamma$ when constructing the waveform.
We require the linear corrections at Newtonian and positive
PN orders to remain smaller than the corresponding GR
contributions at the same PN order. For the negative PN corrections,
which have no GR counterparts, we require them to remain smaller than
the Newtonian contribution throughout the inspiral. The most
restrictive conditions arise from the 1PN correction proportional to
$\alpha$ and the $-1$PN correction proportional to $\gamma$, yielding~\cite{Owen:2025odr}
\begin{equation}
\alpha \, <
\frac{2229+2772\eta}
{1988+14224\eta},
\qquad
\gamma < 7.7\times10^{-2}.
\end{equation}
Over the neutron-star mass range considered in this work, the weakest
of these conditions on $\alpha$ corresponds to a binary with component
masses $m_1=3\,M_\odot$ and $m_2=1\,M_\odot$, giving
$\alpha \lesssim 5.9\times10^{-1}$.
Throughout this work, we restrict the dark-sector parameter space to satisfy the above bounds, thereby ensuring the validity of the perturbative waveform model.

The effect of the chosen dark-sector parameters on the waveform is illustrated in Fig.~\ref{fig:phase_dephasing}, computed over the IndIGO-D observing band, $f_{\rm min}=0.3648\,\mathrm{Hz}$ to $f_{\rm max}=150\,\mathrm{Hz}$, for the same GW170817-like binary at a luminosity distance of $40.4\,\mathrm{Mpc}$. The blue curve denotes the vacuum ($\alpha=\gamma=0$) limit of the waveform, which recovers the Newtonian and 1PN {\IMRPhenomD} terms as noted after Eq.~\eqref{eq:adm_amp_phase}; the orange curve includes the dark-sector corrections, with the phase matched across both activation thresholds as described in Sec.~\ref{subsubsec:matching}.
% As expected from Eq.~\eqref{eq:adm_amp_phase}, the dominant dark-sector contributions enter at negative PN orders and therefore become increasingly important at low frequencies. Since binary NSs spend a significant fraction of their inspiral in the decihertz band, these corrections accumulate coherently over a prolonged observation period, resulting in total dephasings of order \(10^5\)--\(10^6\) rad. This illustrates why observations in the decihertz band provide a particularly sensitive probe of dark-sector physics.
Panel (a) compares the accumulated GW phase predicted by the vacuum and dark-sector waveforms, referenced to zero at $f_{\rm min}$; the aDM curve visibly leads the vacuum curve past the activation thresholds. Panel (b) shows the corresponding dephasing, \(\Delta\Psi=\Psi_{\rm DM}-\Psi_{\rm vac}\). 
% Here $\Psi$ is the accumulated inspiral phase of Eq.~\eqref{eq:TaylorF2_psi}, equivalently the PN series inside $\tilde\varphi(f)$ of Eq.~\eqref{eq:h_series}; the common reference time/phase offset cancels in the difference $\Delta\Psi$ either way.
Even below the activation frequencies, the conservative Yukawa interaction modifies the binary dynamics, leading to a gradual accumulation of dephasing relative to the vacuum waveform. Panel (c) shows the fractional deviation of the Fourier amplitude, $(|\tilde A_{\rm DM}|-|\tilde A_{\rm vac}|)/|\tilde A_{\rm vac}|$, exhibiting piecewise-continuous character of the aDM amplitude, with clean step discontinuities at $f_{\rm quad}$ and $f_{\rm dip}$, as discussed in Sec.~\ref{subsubsec:amplitude_discontinuity}.
% computed from Eq.~\eqref{eq:h_series} removing the common power-law envelope this way makes the piecewise-continuous character of the aDM amplitude unmistakable, with clean step discontinuities at $f_{\rm quad}$ and $f_{\rm dip}$, as discussed in Sec.~\ref{subsubsec:amplitude_discontinuity}. 
The vertical dashed lines in all three panels mark the frequencies at which the electric quadrupole and dipole radiation channels are activated; their activation introduces additional dissipative energy loss, producing distinct changes in the evolution of the accumulated dephasing and the amplitude jumps of panel (c).
Since both activation frequencies lie within the IndIGO-D observing band for the fiducial parameter choice considered here, the detector observes not only the cumulative effect of the conservative Yukawa interaction but also the onset of the dissipative dark-sector radiation channels. The combination of these effects produces large accumulated phase shifts, making the GW phase an especially sensitive probe of the dark-photon mass.

%--------------------------%
\subsection{Methodology}
\label{sec:methodology}
%--------------------------%
%
To investigate the prospects of probing aDM interactions with decihertz GW observations, we perform Bayesian parameter estimation analyses for five representative compact binary systems: two BNS systems with source properties corresponding to GW170817 and GW190425, and three NSBH systems corresponding to GW200105, GW200115 and GW230529. For the BNS analyses, we employ the frequency-domain \texttt{IMRPhenomD\_NRTidalv2} waveform model, while the NSBH systems are analyzed using the \texttt{IMRPhenomNSBH}~\cite{Thompson:2020nei} waveform model. In each case, we incorporate the amplitude and phase correction described in Sec.~II, parameterized by the dark-sector parameters $\alpha$, $\gamma$, and $\mu$, into the corresponding waveform model.

The source parameters are inferred within the Bayesian framework, where the posterior probability distribution is obtained by combining the likelihood with the prior according to Bayes' theorem,
\begin{equation}
    p(\vec{\Lambda}\mid d)=
    \frac{\mathcal{L}(d\mid\vec{\Lambda})\,\pi(\vec{\Lambda})}
         {p(d)},
\end{equation}
where $\vec{\Lambda}$ denotes the complete set of source parameters, $\mathcal{L}(d\mid\vec{\Lambda})$ is the likelihood, $\pi(\vec{\Lambda})$ is the prior, and $p(d)$ is the Bayesian evidence. We utilize the phase-marginalized log-likelihood function, which under the assumption of stationary Gaussian detector noise is given by~\cite{Thrane:2018qnx},
\begin{equation}
    \ln \mathcal{L}(d\mid\vec{\Lambda}) = \ln I_0 \left| 
    \langle h(\vec{\Lambda}) \mid d \rangle_{\mathbb{C}}
    \right| -
    \frac{1}{2} \left\| h(\vec{\Lambda}) \right\|^2,
    \label{eq:logl_basic}
\end{equation}
where $I_0$ is the zeroth-order modified Bessel function of the first kind, and
\begin{equation}
    \langle a|b\rangle
    = % {\rm Re} \, \langle a|b\rangle_{\mathbb{C}} = 
    4\,{\rm Re}
    \int_{f_{\rm min}}^{f_{\rm max}}
    \frac{\tilde a^{*}(f)\tilde b(f)}
         {S_n(f)}
    \,df
    \label{eq:inner_product}
\end{equation}
is the standard noise-weighted inner product, with $S_n(f)$ denoting the one-sided detector noise power spectral density; $\fmin$ and $\fmax$ are the lower and upper cut-off frequencies; ${\left\Vert a \right\Vert^2 = \inp{a}{a}}$ denotes the norm and $\tilde{a}(f)$ denotes the Fourier transform of $a(t)$. The argument of $\ln I_0$ in Eq.~\eqref{eq:logl_basic} is a complex inner-product, not just the real part of the integral as in Eq.~\eqref{eq:inner_product}.

\begin{figure}
    \centering
    \includegraphics[width=1\linewidth]{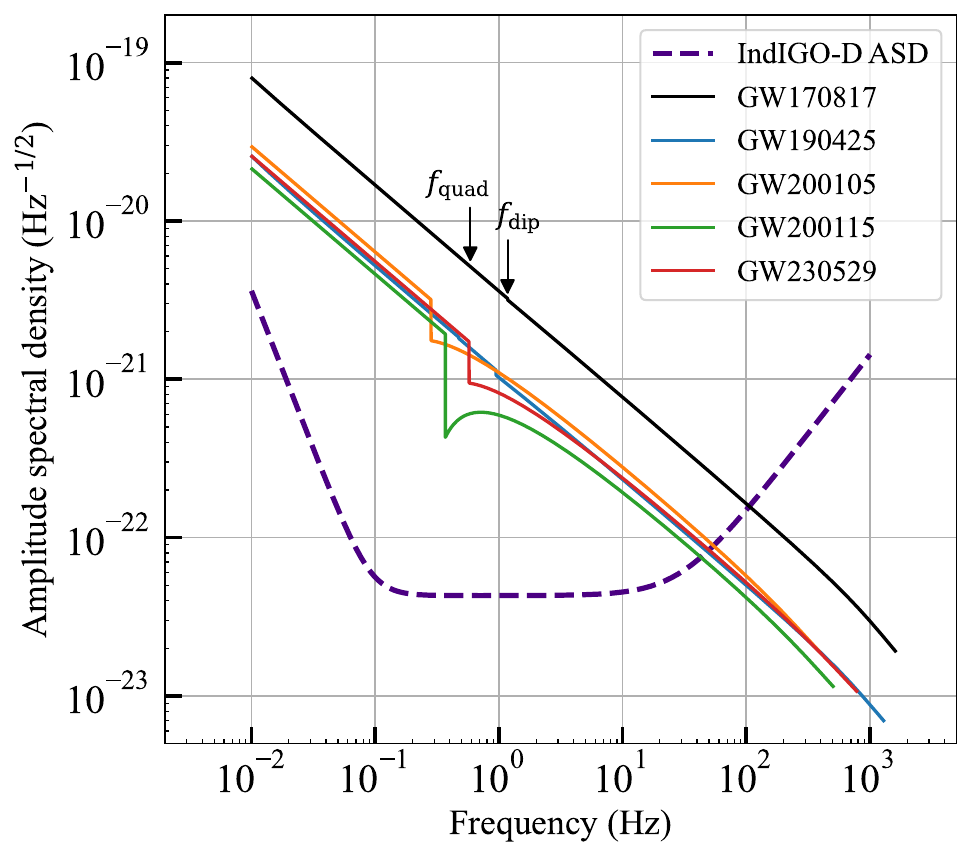}
    % \caption{Amplitude spectral densities of the GW signals for GW170817, GW190425, GW200105, GW200115, and GW230529-like events plotted up to the frequency corresponding to the innermost stable circular orbit, shown together with the IndIGO-D noise amplitude spectral density (purple). The signals are computed incorporating the effects due to aDM. The arrows mark the onset of dark quadrupole and dipole radiations, \(f_{\rm quad}\) and \(f_{\rm dip}\), respectively, for a GW170817-like event.}
    \caption{Amplitude spectral densities of the GW signals for GW170817, GW190425, GW200105, GW200115, and GW230529-like events, plotted up to the frequency corresponding to the innermost stable circular orbit, together with the IndIGO-D noise amplitude spectral density (purple)~\cite{IndIGOD_repo}. The signals are computed including the effects of aDM. The arrows mark the frequencies corresponding to the onset of dark quadrupole and dipole radiation, $f_{\rm quad}$ and $f_{\rm dip}$, respectively, for the GW170817-like system. The discontinuities in the signal amplitudes arise from the opening of these additional dark-radiation channels once the corresponding threshold frequencies are reached.}
    \label{fig:asd}
\end{figure}

% The IndIGO-D amplitude spectral density (ASD) adopted in this work is shown in Fig.~\ref{fig:asd}. 
We adopt the fiducial one-sided IndIGO-D noise power spectral density $S_n(f)$ provided by the IndIGO-D Collaboration~\cite{IndIGOD_repo}. Its corresponding amplitude spectral density (ASD), $\sqrt{S_n(f)}$, is shown in Fig.~\ref{fig:asd}.
The figure also shows the ASDs of representative GW170817-, GW190425-, GW200105-, GW200115-, and GW230529-like signals, computed including the aDM corrections and plotted up to the frequency corresponding to the innermost stable circular orbit. The values of the DM parameters used for generating these waveforms are given in Table~\ref{tab:pe_summary}. The arrows labeled \(f_{\rm quad}\) and \(f_{\rm dip}\) denote the electric quadrupole and dipole activation frequencies, respectively, for the GW170817-like system. For the fiducial dark-sector parameters adopted in this work, both activation frequencies lie within the sensitive decihertz band of IndIGO-D, allowing the detector to observe the onset of the additional dark-radiation channels.

For all analyses, the upper frequency cutoff is fixed to $f_{\rm max}=10~{\rm Hz}$, while the lower cutoff frequency is chosen such that the in-band duration of the signal is three months.
The parameter space consists of the intrinsic binary parameters, $\vec{\Lambda}_{\rm{int}}$, including the component masses, spins, together with the dark-sector parameters $(\alpha,\gamma,\mu)$, and the extrinsic parameters, $\vec{\Lambda}_{\rm{ext}}$, describing the source orientation, sky location, luminosity distance, polarization, and coalescence time. 
The detector-frame strain in the frequency domain can be written as
a linear combination of the two GW polarizations,
%And, $h(\vec{\Lambda}, t)$ is the induced strain in the detector that can be written as a linear combination of the GW polarization amplitudes $h_{+,\times}(t)$:
%
\begin{equation}
\tilde h_d(f)
=
e^{-2\pi i f\,\hat{k}\cdot\vec{x}_d[t(f)]/c}
\sum_{a=+,\times}
F_a\!\left(\hat{k},\psi,t(f)\right)
\tilde h_a(f).
%h(\vec{\Lambda}, t) = e^{-2\pi i f\,\hat{k}\cdot\vec{x}_d(t)/c} \sum_{a=+,\times} F_a(\hat{k},\psi,t)\,h_a(\vec{\Lambda}_{\rm{int}}, t).
\label{eq:detector_response}
\end{equation}
where, $\vec{x}_d(t)$ denotes the position of the detector and $\hat{k}$ denotes the direction of GW propagation measured in the solar-system barycentric (SSB) frame, and $F_{+,\times}$ are the antenna pattern functions corresponding to the two polarization states. The exponential factor accounts for the propagation-time delay between the detector and the SSB. For the L-shaped heliocentric IndIGO-D configuration, the time-dependent detector position and antenna pattern functions are computed following Ref.~\cite{Sharma:2026tff}. 
For long-duration signals, detector motion causes this delay to vary during signal accumulation, producing Doppler phase modulation of the observed signal. The orbital motion also causes the antenna pattern functions to vary with time, leading to a characteristic amplitude modulation of the observed signal. These Doppler and antenna-pattern modulations accumulate over the observation period and carry valuable information about the source location. 
% For the L-shaped heliocentric IndIGO-D configuration, the detector's orbital position and the full frequency-dependent antenna response are computed following Ref.~\cite{Sharma:2026tff}. 

% Since IndIGO-D follows a heliocentric orbit, the detector response varies continuously during the observation due to the orbital motion of the constellation. These time-dependent modulation effects are incorporated through the corresponding antenna pattern functions and detector time delays. 
%
\begin{figure*}[!hbt]
    \centering
    \includegraphics[width=0.75\linewidth]{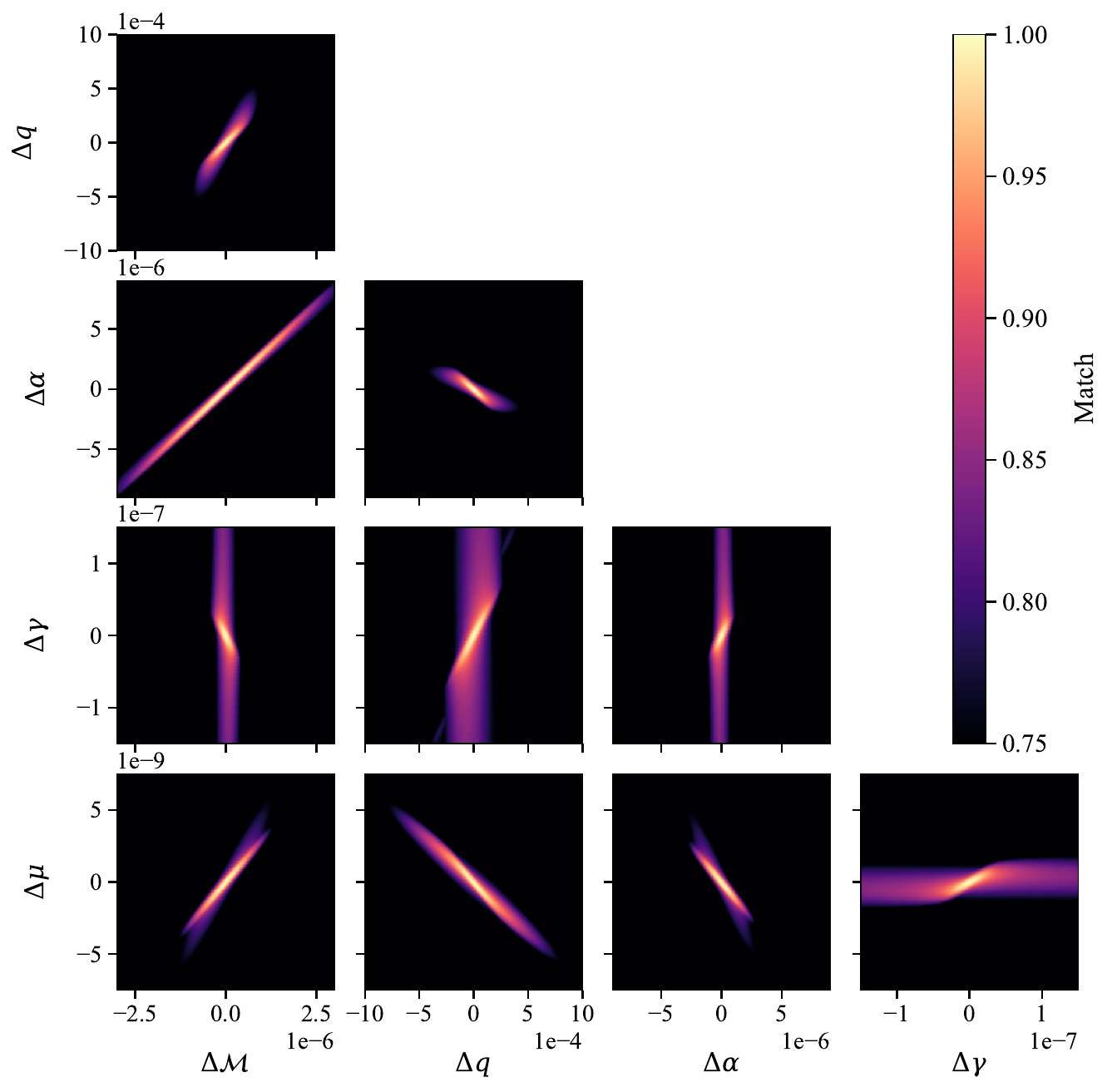}
    \caption{Match surfaces for all ten orthogonal two-dimensional slices through the five-dimensional intrinsic parameter space $(\mathcal{M}, q, \alpha, \gamma, \mu)$, centered on the GW170817 injected parameters. In each panel, only the two displayed parameters are varied while the remaining three are held fixed at their injected values. The injected waveform is located at the origin of each panel ($\Delta\Lambda_{\rm int}=0$), and the color denotes the match between the injected waveform and neighboring templates. These local match surfaces are used to estimate suitable prior widths for the intrinsic parameters and to identify the region over which the relative binning approximation remains accurate.}
    \label{fig:match_surfaces}
\end{figure*}

As the parameter estimation is carried out entirely in the frequency domain, the detector response is likewise evaluated in the frequency domain by exploiting the SPA.  
The time-frequency correspondence is obtained from
\begin{equation}
t(f) = \frac{1}{2\pi}
\frac{d\tilde{\varphi}(f)}{df},
\label{eq:time_frequency}
\end{equation}
where $\tilde{\varphi}(f)$ is the total Fourier-domain phase defined
in Eq.~\eqref{eq:h_series_phase}. For a given set of intrinsic binary
parameters, Eq.~\eqref{eq:time_frequency} maps each instantaneous
GW frequency to the corresponding epoch during the
inspiral.
The time-dependent antenna pattern functions are then evaluated at $t(f)$, allowing the detector response to be constructed directly in the frequency domain while fully accounting for the orbital motion of the detector.

The parameter estimation analyses are performed using the \texttt{PyCBC Inference}~\cite{Biwer:2018osg} library. Bayesian parameter estimation for compact binary coalescences requires evaluating the likelihood function $\sim10^{7}$--$10^{8}$ times over the source parameter space. Since each likelihood evaluation involves generating detector-frame waveforms and computing noise-weighted inner products, a direct implementation rapidly becomes computationally prohibitive, particularly for long-duration signals observed by IndIGO-D. To accelerate the likelihood evaluation, we employ the relative binning method~\cite{Cornish:2010kf, Cornish:2021lje, Zackay:2018qdy} and use the true injection parameters as the fiducial reference waveform parameters. Posterior samples are generated using the \texttt{dynesty}~\cite{Speagle:2019ivv} nested sampler through the \texttt{Bilby}~\cite{Ashton:2018jfp} interface with the \texttt{acceptance-walk} sampling method, adopting ${\tt nlive}=3000$ and ${\tt naccept}=60$.

%--------------------------------------------------------%
\subsection{Exploration of the intrinsic parameter space}
\label{subsec:intrinsic_ps_match}
%--------------------------------------------------------%
%
The accuracy of the relative binning approximation relies on the proposed waveform remaining sufficiently close to the fiducial waveform used to construct the frequency bins. Significant deviations from the fiducial waveform can lead to a degradation of the approximation and consequently reduce the accuracy of the likelihood evaluation. It is therefore advantageous to identify, prior to stochastic sampling, the region of the intrinsic parameter space that is expected to contain the dominant posterior support. To this end, we investigate the local structure of the likelihood using the match function between neighboring waveforms. Since the match quantifies the similarity between two normalized waveforms, it provides an efficient approximation to the local geometry of the likelihood. Since the treatment for all systems considered in this work is identical, we illustrate the procedure using the GW170817-like binary. The match between two normalized waveforms $a$ and $b$ is defined as
\begin{equation}
\mathcal{M}(a,b)
=
\max_{t_{\rm ref},\,\varphi_{\rm ref}} \langle \hat a \mid \hat b \rangle,
\label{eq:Match}
\end{equation}
where $\hat{a} = a/\sqrt{\inp{a}{a}}$ denotes the normalized waveform, and the maximization is performed over an arbitrary coalescence time and reference phase. 

Figure~\ref{fig:match_surfaces} shows the orthogonal two-dimensional match surfaces for all parameter pairs in the five-dimensional intrinsic parameter space $(\mathcal{M},q,\alpha,\gamma,\mu)$, with each slice centered on the injected parameters. The extent of the high-match region varies significantly between different parameter directions, indicating that the waveform is substantially more sensitive to perturbations in some parameters than in others. These match surfaces therefore provide a useful estimate of the local scales over which the likelihood remains appreciable around the injected waveform. It should be emphasized that these match surfaces represent local two-dimensional slices through the likelihood with all remaining intrinsic parameters fixed. They should therefore not be interpreted as the marginalized posterior distributions obtained from Bayesian inference, in which all intrinsic and extrinsic parameters are varied simultaneously. Nevertheless, they provide a useful guide to the characteristic scales of the likelihood around the fiducial waveform, which is the quantity relevant for selecting appropriate prior widths and ensuring the validity of the relative binning approximation.

Although the match is not identical to the posterior distribution, the two are closely related. The log-likelihood decreases approximately in proportion to the waveform mismatch, with the suppression becoming increasingly severe as the signal-to-noise ratio (SNR) increases. Since all injections considered in this work have SNRs exceeding $30$, with the GW170817-like system having an SNR of approximately $80$, the posterior support is expected to be confined to a very high-match region surrounding the injected parameters. Figure~\ref{fig:match_surfaces} displays match values down to a threshold of $0.75$, substantially below the region that contributes appreciably to the posterior. Consequently, the widths inferred from these local match surfaces provide a practical estimate of the region that is expected to contain the dominant posterior support.

However, the match surfaces do not capture correlations between the intrinsic and extrinsic source parameters, nor do they represent the exact likelihood obtained after marginalizing over the full parameter space. To ensure that no physically relevant posterior support is excluded, we therefore adopt prior ranges that are considerably broader than those suggested by the local match structure. 
% In practice, the prior widths for all intrinsic parameters are chosen to be approximately two orders of magnitude larger than the widths indicated by the match surfaces. 
%This choice preserves the accuracy of the relative binning approximation while allowing the sampler to efficiently explore the complete region of significant posterior support.
This choice preserves the accuracy of the relative-binning approximation while allowing the sampler to explore the range of parameters that give appreciable likelihood around the injected signal. The resulting credible intervals should therefore be interpreted as estimates of how precisely the dark-sector parameters could be measured if the signal is close to the benchmark values considered here.
\begin{table}[!t]
    \def\arraystretch{1.25}
    \centering
    \begin{tabular}{l l l}
    \toprule[1pt]
        Parameters     &\:\:  Range    &   Prior distribution \\    
        \midrule[1pt]
        $\mathcal{M}$  &   $[\mchirp_{\rm{inj}} \pm 10^{-3}~\msun]$           &   $\propto \mathcal{M}$ \\
        $q$            &   $[1, \;q_{\rm{inj}} + 1]$            & $ \propto \left [ (1 + q)/q^3 \right ]^{2/5}$ \\
        $\chi_{1z}$ (BNS)  &   $[-0.05,\; 0.05]$    & Uniform  \\
        $\chi_{1z}$ (NSBH)       &   $[-0.5,\; 0.5]$      & Uniform  \\
        $\chi_{2z}$       &   $[-0.05,\; 0.05]$    & Uniform  \\
        $\dL$          & $[1,\;1000] \Mpc$            & Uniform in volume\\
        $t_c$          & $[t_c^{\rm{inj}} \pm 30~\rm{s}]$ & Uniform\\
        $\lambda$       & $[0, 2\pi]$            & Uniform\\
        $\beta$       & $[-\pi/2, \pi/2]$        & $\sin^{-1} \left [ {\text{Uniform}}[-1,1]\right ]$\\
        $\iota$        & $[0, \pi]$             & Uniform in $\cos \iota$\\
        \midrule
        $\alpha$       & $[\alpha_{\rm{inj}} \pm \delta\alpha]$ & Uniform \\
        $\gamma$       & $[\gamma_{\rm{inj}} \pm \delta\gamma]$ & Uniform \\
        $\mu$          & $[\mu_{\rm{inj}} \pm \delta\mu]$       & Uniform \\
    \bottomrule[1pt]
    \end{tabular}
    \caption{The prior distributions over various parameters and their corresponding ranges used in the parameter estimation with IndIGO-D. The chosen prior distributions over $\mchirp$ and $q$ correspond to uniform distributions over component masses. Priors over ecliptic longitude $\lambda$ and latitude $\beta$ correspond to an isotropic distribution across the sky.
    % For the aligned spin components, narrower priors are adopted for neutron stars ($\chi_z \in [-0.05, 0.05]$) compared to the primary black hole in NSBH systems ($\chi_{1z} \in [-0.5, 0.5]$).
    For the dark-sector parameters ($\alpha, \gamma, \mu$), uniform priors are centered on the injected values with widths $\delta\theta$ listed in Table~\ref{tab:pe_summary}.}
    \label{tab:priors}
\end{table}
\begin{table*}
    \centering
    \resizebox{\textwidth}{!}{%
    \renewcommand{\arraystretch}{1.65}
    \begin{tabular}{l
                    ccc
                    cc
                    ccc
                    ccc}
        \toprule
        Event
        & $\alpha_{\rm inj}$
        & $\gamma_{\rm inj}$
        & $\mu_{\rm inj}$
        & $f_{\min}\,[{\rm Hz}]$
        & $\rho_{\rm opt}$
        & $\delta\alpha$
        & $\delta\gamma$
        & $\delta\mu$
        & $\alpha$
        & $\gamma$
        & $\mu$
        % & $m_Ac^2\,[{\rm eV}]$
        \\
        \midrule
        \multicolumn{12}{c}{Binary neutron star systems}
        \\
        \midrule
        GW170817
        & $0.035$
        & $4.1\times10^{-4}$
        & $5.0\times10^{-5}$
        & $0.364$
        & $77.9$
        & $1.5\times10^{-4}$
        & $5.0\times10^{-7}$
        & $1.5\times10^{-8}$
        & $0.035^{+3.02\times10^{-5}}_{-2.97\times10^{-5}}$
        & $\left(4.1^{+1.37\times10^{-3}}_{-1.34\times10^{-3}}\right)
           \times10^{-4}$
        & $\left(5.0^{+1.55\times10^{-4}}_{-1.52\times10^{-4}}\right)
           \times10^{-5}$
        % & $\left(2.42^{+2.59\times 10^{-5}}_{-2.63 \times 10^{-5}}\right)\times10^{-15}$
        \\
        GW190425
        & $0.11$
        & $1.5\times10^{-3}$
        & $5.0\times10^{-5}$
        & $0.318$
        & $33.8$
        & $5.0\times10^{-3}$
        & $1.0\times10^{-6}$
        & $1.0\times10^{-7}$
        & $0.11^{+5.58\times10^{-5}}_{-5.83\times10^{-5}}$
        & $\left(1.5^{+3.10\times10^{-4}}_{-3.18\times10^{-4}}\right)
           \times10^{-3}$
        & $\left(5.0^{+1.48\times10^{-4}}_{-1.51\times10^{-4}}\right)
           \times10^{-5}$
        % & $\left(1.95^{+1.15\times 10^{-5} }_{-1.23\times 10^{-5}}\right)\times10^{-15}$
        \\
        \midrule
        \multicolumn{12}{c}{Neutron star--black hole systems}
        \\
        \midrule
        GW200105
        & $0$
        & $1.0\times10^{-2}$
        & $5.0\times10^{-5}$
        & $0.182$
        & $37.3$
        & ---
        & $5.0\times10^{-4}$
        & $5.0\times10^{-6}$
        & fixed
        & $\left(9.99^{+1.73\times10^{-2}}_{-2.09\times10^{-2}}\right)
           \times10^{-3}$
        & $\left(4.99^{+0.013}_{-0.0157}\right)\times10^{-5}$
        % & $\left(5.83^{+1.05 \times 10^{-5}}_{-1.14\times 10^{-5}}\right)\times10^{-16}$
        \\
        GW200115
        & $0$
        & $1.7\times10^{-2}$
        & $5.0\times10^{-5}$
        & $0.225$
        & $38.1$
        & ---
        & $5.0\times10^{-4}$
        & $5.0\times10^{-6}$
        & fixed
        & $0.017^{+1.84\times10^{-5}}_{-2.11\times10^{-5}}$
        & $\left(5.0^{+0.00812}_{-0.00931}\right)\times10^{-5}$
        % & $\left(7.63^{+1.28\times 10^{-5}}_{1.29\times 10^{-5}}\right)\times10^{-16}$
        \\
        GW230529
        & $0$
        & $1.0\times10^{-2}$
        & $5.0\times10^{-5}$
        & $0.262$
        & $37.2$
        & ---
        & $5.0\times10^{-4}$
        & $1.0\times10^{-5}$
        & fixed
        & $0.01^{+2.02\times10^{-5}}_{-1.52\times10^{-5}}$
        & $\left(5.01^{+0.0151}_{-0.0114}\right)\times10^{-5}$
        % & $\cdots$
        \\
        \bottomrule
    \end{tabular}%
    }
    \caption{Summary of Bayesian parameter estimation setup and the recovered dark-sector parameters for the simulated BNS and NSBH events in \text{IndIGO-D}. Other non-DM parameters are set to the MAP values from the LVK catalog data release, while
    %the dark-sector parameters ($\alpha, \gamma, \mu$) are set to the 90\% (credible level) CL upper bounds from Owen~\textit{et al.}~\cite{Owen:2025odr}. 
    the injected $\alpha$ and $\gamma$ values are benchmark values taken
    from the 90\% credible upper limits reported by Owen~\textit{et al.}
    The lower cutoff frequency for each binary $f_{\text{min}}$ corresponds to a three-month in-band observation period in IndIGO-D. The quantities $\delta\alpha$, $\delta\gamma$, and $\delta\mu$ are the half-widths of the uniform priors centered on the injected values; a dash indicates a parameter that is fixed rather than sampled. The recovered values are posterior medians with $90\%$ credible intervals bounded by the $5^{\rm th}$ and $95^{\rm th}$ percentiles. The quoted posteriors are obtained from the zero-noise analyses.}
    \label{tab:pe_summary}
\end{table*}
In summary, Figure~\ref{fig:match_surfaces} identifies the characteristic scales over which the waveform remains nearly indistinguishable from the injected signal, thereby providing an estimate of the region expected to contain the dominant posterior support. Second, they guide the choice of prior widths for the intrinsic parameters, ensuring that the sampler explores all physically relevant regions without venturing unnecessarily far from the fiducial waveform, where the relative binning approximation gradually loses accuracy. Although the final Bayesian inference is performed in the full multidimensional parameter space, the match surfaces provide a simple and computationally efficient tool for constructing robust and well-motivated priors.
%
%%=======================%%
\section{Results}
\label{sec:results}
%%=======================%%
%
We now present the results of Bayesian parameter estimation on simulated signals for five representative systems observed in the decihertz band with the proposed \text{IndIGO-D} observatory. Our target sources comprise two BNS events, analogous to GW170817 and GW190425, and three NSBH events, analogous to GW200105, GW200115, and GW230529. 
For all simulated events, we adopt benchmark values of the dark-sector
parameters ($\alpha$ and $\gamma$) corresponding to the 90\% credible
upper limits reported in the dedicated aDM analysis of
Owen~\textit{et al.}~\cite{Owen:2025odr}.
These benchmark values are used consistently for the forecasts and
should not be interpreted as the strongest current bounds; the
independent constraints inferred from parameterized tests of GR are
discussed in Sec.~\ref{subsec:adm_past_gw}.
%
%For all simulated events, the injected dark-sector parameters ($\alpha$ and $\gamma$) are set to the 90\% credible level (CL) upper limits established by Owen~\textit{et al.}~\cite{Owen:2025odr} from ground-based detector analyses. 
The analysis of Owen~\textit{et al.} in the ground-based detector band
does not meaningfully constrain $\mu$. We therefore adopt a fiducial
value $\mu = 5 \times 10^{-5}$ for all simulated events.

For NSBH systems, we assume that the black hole carries no dark
electric charge. The conservative Yukawa coupling therefore vanishes,
and we set $\alpha=0$ for these sources.
The remaining source parameters are fixed to the maximum
\textit{a posteriori} values of the LVK posterior estimates from the
corresponding gravitational-wave transient catalog releases
\cite{LIGOScientific:2018hze, GW170817_posterior_samples,
gwtc_2.1, GWTC2p1_PE_data, gwtc_3, GWTC3_PE_data}.
Note that the parameters describing the position of the source and coalescence time are transformed from geocentric estimates into the SSB frame. The injection DM parameters, lower cut-off frequency, the optimal SNR, half-widths of the uniform dark-sector priors ($\delta\alpha, \delta\gamma, \delta\mu$), and median value of the recovered dark-sector parameter with 90\% credible interval (CI) for all the systems are summarized in Table~\ref{tab:pe_summary}. 
The prior distributions for various parameters and their respective ranges used while sampling the posterior distribution in all analyses are summarized in Table~\ref{tab:priors}.
The dark-sector priors are centered on the benchmark injected values,
as motivated by the local match analysis of Sec.~\ref{subsec:intrinsic_ps_match}.
Accordingly, the credible intervals quoted in Table~\ref{tab:pe_summary} characterize the achievable parameter recovery precision in the vicinity of these benchmark signals.

\begin{figure*}[!t]
    \centering
    \includegraphics[width=\textwidth]
    {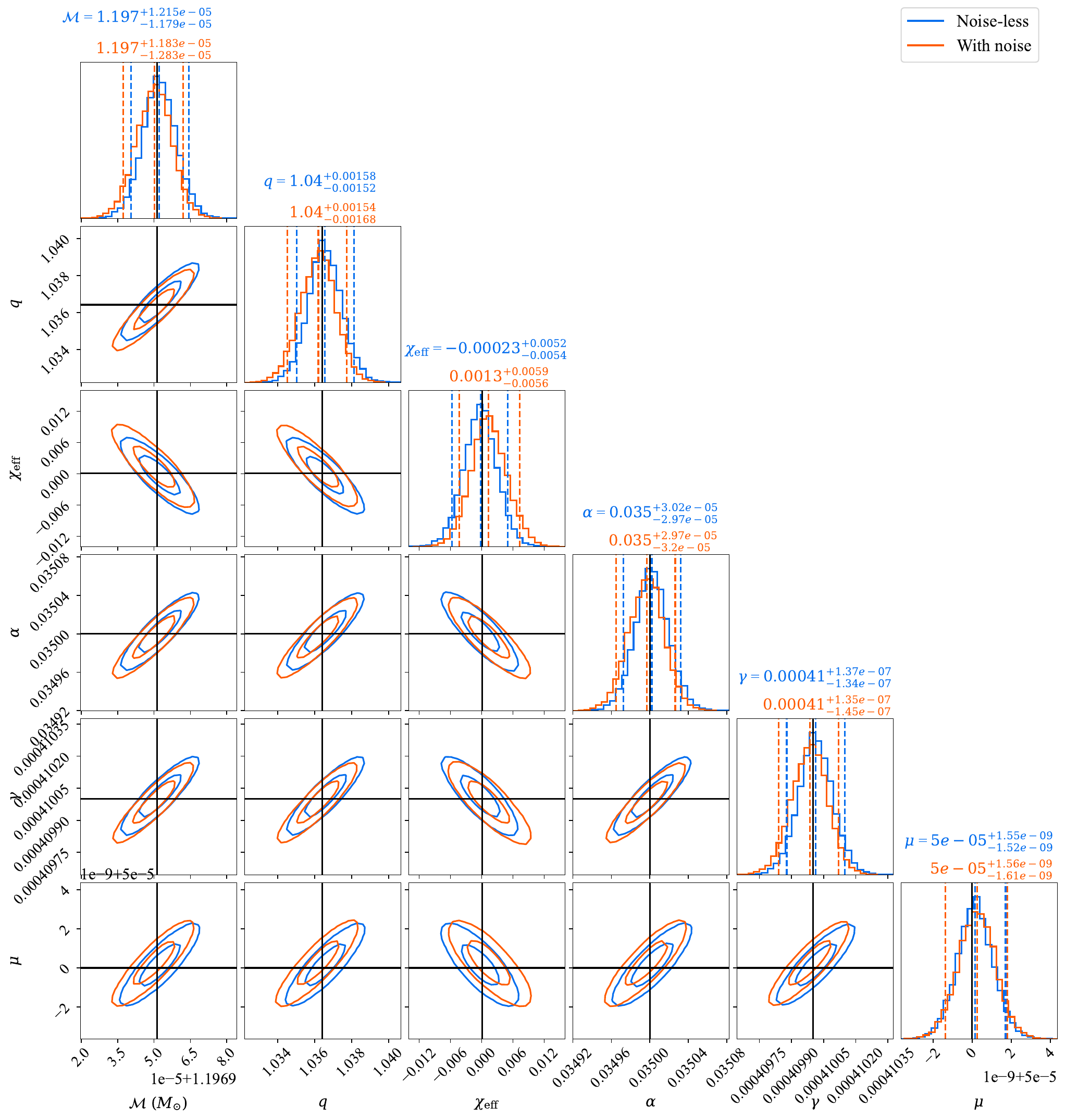}
    \caption{
    Comparison of the inferred marginalized intrinsic posterior distributions for the GW170817-like BNS system obtained from simulated
    zero-noise data (blue) and from data containing a realization of stationary Gaussian noise colored by the IndIGO-D PSD (orange). The displayed parameters are the detector-frame chirp mass $\mathcal{M}$, mass ratio $q$, effective aligned spin
    $\chi_{\rm eff}$, Yukawa coupling strength $\alpha$, dipole parameter $\gamma$, and mediator-mass parameter $\mu$. The contours in the marginalized 2-dimensional distributions correspond to 50\% and 90\% credible levels, and solid black lines denote the injected values. The labels above each column represent the median and 90\% CI for the corresponding parameter. The close overlap of the two posterior distributions demonstrates that the intrinsic and dark-sector parameter measurements are stable against the inclusion of a representative Gaussian noise realization.
    }
    \label{fig:gw170817_intrinsic_comparison}
\end{figure*}

\begin{figure*}[t!]
    \centering
    \includegraphics[width=0.8\textwidth]
    {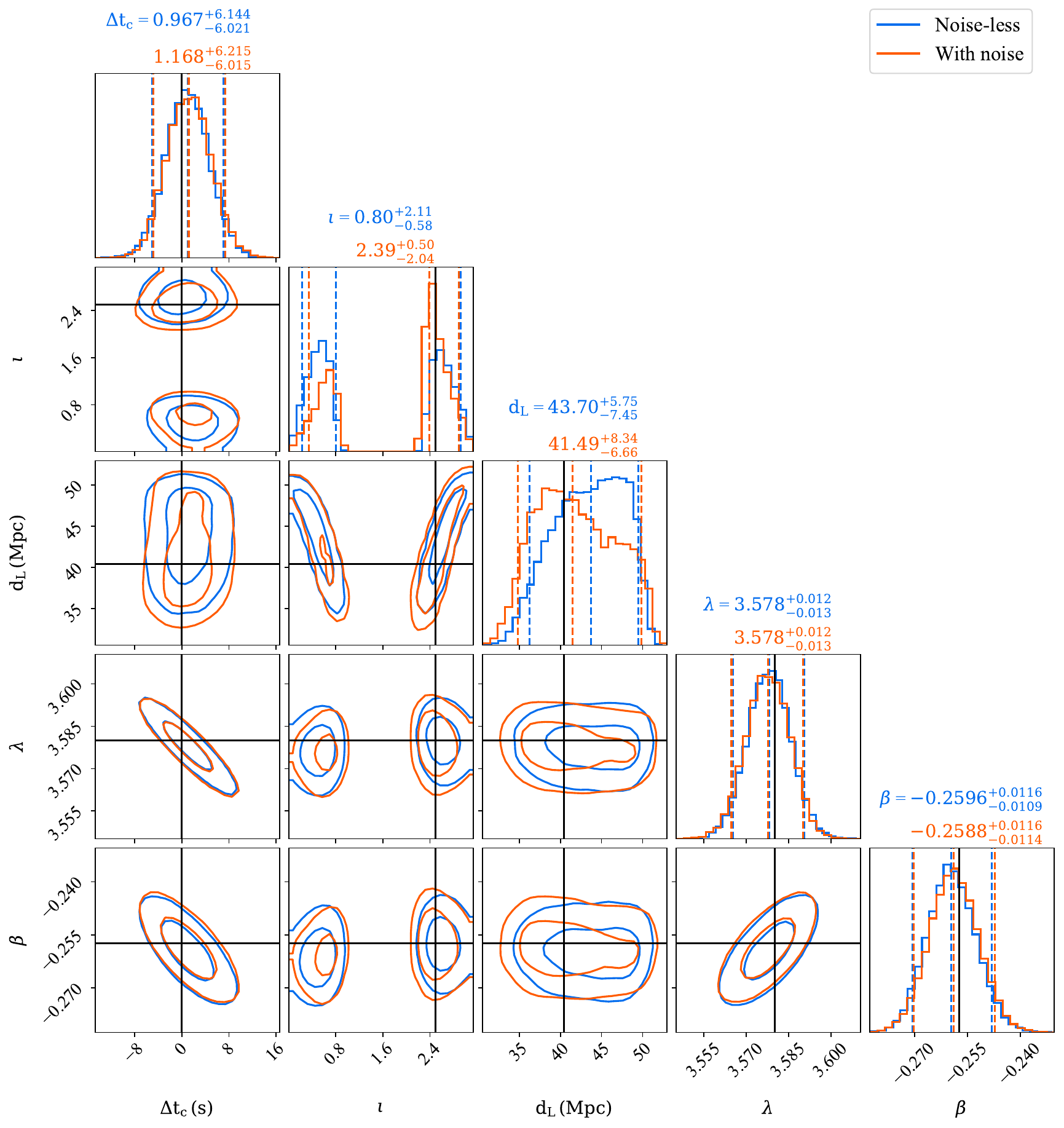}
    \caption{
    Comparison of the inferred marginalized extrinsic posterior distributions for the GW170817-like system obtained using zero-noise simulated data (blue) and data containing stationary Gaussian noise colored by the IndIGO-D PSD (orange). The parameters shown are the coalescence-time offset $\Delta t_c$ (in SSB frame), inclination angle $\iota$, luminosity distance $d_L$, ecliptic longitude $\lambda$, and ecliptic latitude $\beta$. The contours in the marginalized 2-dimensional distributions correspond to 50\% and 90\% credible levels, and solid black lines denote the injected values. The labels above each column represent the median and 90\% CI for the corresponding parameter. The comparable posterior widths in the two analyses demonstrate that the recovery of the extrinsic parameters remains stable after the inclusion of a representative Gaussian noise realization.
    }
    \label{fig:gw170817_extrinsic_comparison}
\end{figure*}

For the GW170817-like system, we perform two separate parameter-estimation analyses. In the first, the injected signal is analyzed without adding a particular realization of detector noise, while in the second the same signal is embedded in stationary Gaussian noise colored by the IndIGO-D noise PSD. 
%The zero-noise analysis provides a representative estimate of the expected measurement precision, corresponding to the likelihood averaged over an ensemble of noise realizations. 
% The zero-noise analysis provides a representative estimate of the expected measurement precision, free from fluctuations associated with a particular noise realization. 
The analysis including one explicit
Gaussian-noise realization provides a complementary
test of the robustness of the inference against
statistical fluctuations in the detector data.
Figures~\ref{fig:gw170817_intrinsic_comparison} and~\ref{fig:gw170817_extrinsic_comparison} compare the marginalized posterior distributions in intrinsic and extrinsic parameters, respectively, obtained from the zero-noise and noisy analyses for the GW170817-like system. The two sets of posteriors show very similar widths and correlation structures, with only small noise-induced shifts in their medians. In the remainder of this section, we therefore quote results from the zero-noise analyses, which provide a clean forecast of the expected measurement precision without fluctuations associated with a particular noise realization. 
In the zero-noise analysis, the dark-sector parameters are recovered as
\begin{align}
    \nonumber
    \alpha
    &=
    \left(\alpha_{\rm inj} + 4.42 \times 10^{-6}\right)^{+3.02\times10^{-5}}_{-2.97\times10^{-5}},
    \\  \nonumber
    \gamma
    &=
    \left(\gamma_{\rm inj} + 2.26 \times 10^{-8}\right)^{+1.37\times10^{-7}}_{-1.34\times10^{-7}},
    \\  \nonumber
    \mu
    &=
    \left(\mu_{\rm inj} + 2.55 \times 10^{-10}\right)^{+1.55\times 10^{-9}}_{-1.52\times10^{-9}},
\end{align}
where the quoted bounds denote the median and the  $90\%$ credible interval bounded by the $5^{\rm th}$ and $95^{\rm th}$ percentiles. The posterior median is written as the sum of the injected value and the corresponding offset. The corresponding noisy analysis gives similar values. The close agreement between the two analyses demonstrates that the measurement of the dark-sector parameters is stable against the inclusion of a representative Gaussian-noise realization. The conventional binary parameters are also measured with high precision. For the zero-noise analysis, the chirp mass and mass ratio are recovered as
\begin{align}
    \nonumber
    \mathcal{M}
    &=
    \left(\mathcal{M}_{\rm inj} + 9.41 \times 10^{-7}\right)^{+1.21\times10^{-5}}_{-1.18\times10^{-5}}
    \,M_{\odot},
    \\ \nonumber
    q
    &=
    \left(q_{\rm inj} + 1.26 \times 10^{-4}\right)^{+0.00158}_{-0.00152}.
\end{align}
\begin{figure*}[!t]
    \centering
    \includegraphics[width=0.8\textwidth]
    {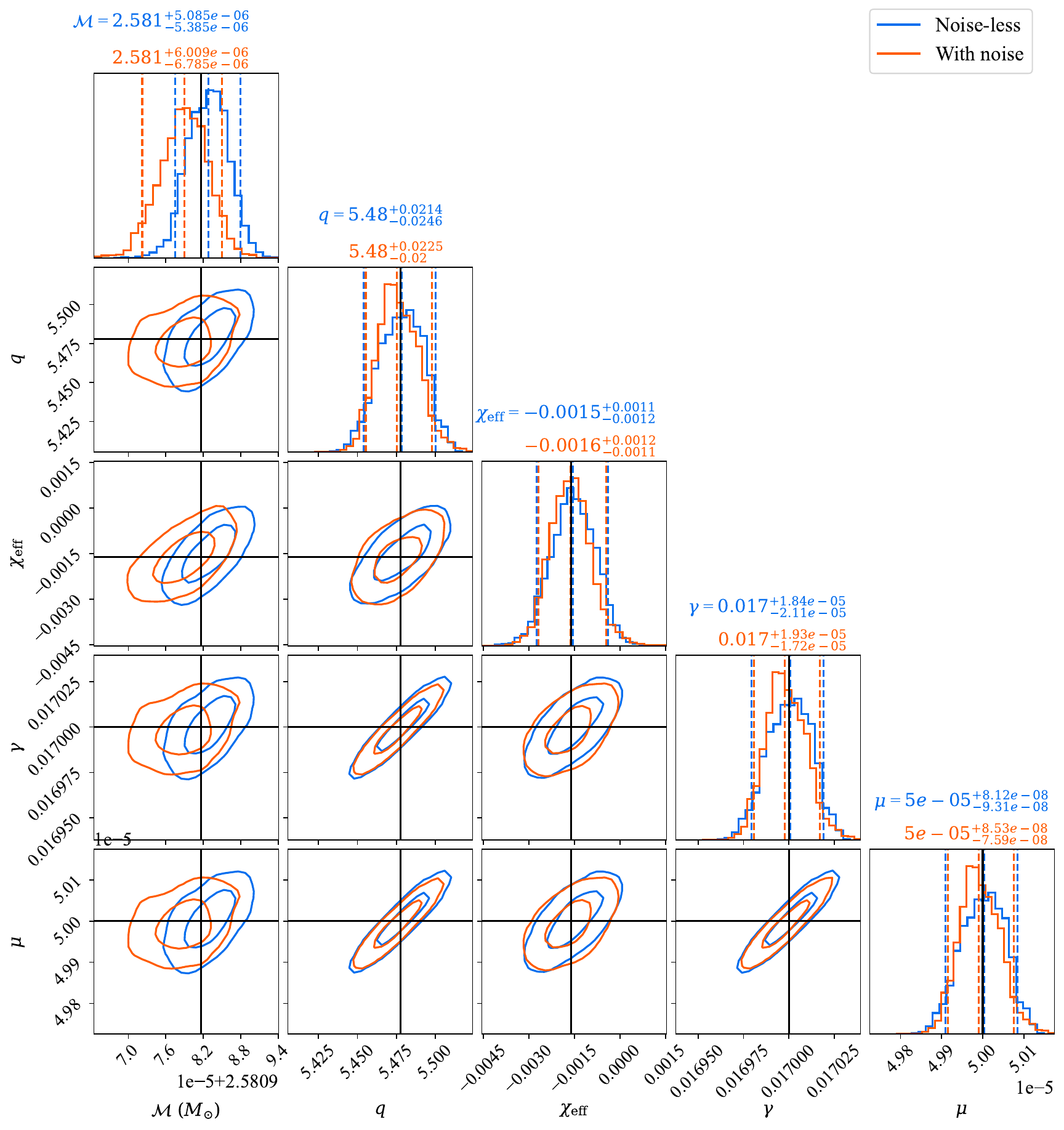}
    \caption{
    Comparison of the inferred marginalized intrinsic posterior distributions for the GW200115-like NSBH system obtained from simulated
    zero-noise data (blue) and from data containing a realization of stationary Gaussian noise colored by the IndIGO-D PSD (orange). The displayed parameters are the detector-frame chirp mass $\mathcal{M}$, mass ratio $q$, effective aligned spin
    $\chi_{\rm eff}$, dipole parameter $\gamma$, and mediator-mass parameter $\mu$. The contours in the marginalized 2-dimensional distributions correspond to 50\% and 90\% credible levels, and solid black lines denote the injected values. The labels above each column represent the median and 90\% CI for the corresponding parameter. The close overlap of the two posterior distributions demonstrates that the intrinsic and dark-sector parameter measurements are stable against the inclusion of a representative Gaussian noise realization.
    }
    \label{fig:gw200115_intrinsic_comparison}
\end{figure*}

For a GW190425-like BNS system, the recovered median values and 90\% CI bounds for the intrinsic parameters are,
\begin{align}
\nonumber
    \mathcal{M}
    &=
    \left(\mathcal{M}_{\rm inj} + 2.15 \times 10^{-6}\right)^{+2.77\times10^{-5}}_{-2.89\times10^{-5}}
    \,M_{\odot},
    \\ \nonumber
    q
    &=
    \left(q_{\rm inj} + 5.01 \times 10^{-4}\right)^{+9.25\times10^{-4}}_{-9.55\times10^{-4}},\\
\nonumber
    \alpha
    &=
    \left(\alpha_{\rm inj} + 4.18 \times 10^{-6}\right)^{+5.58\times10^{-5}}_{-5.83\times10^{-5}},
    \\  \nonumber
    \gamma
    &=
    \left(\gamma_{\rm inj} + 1.77 \times 10^{-8}\right)^{+3.10\times10^{-7}}_{-3.18\times10^{-7}},
    \\  \nonumber
    \mu
    &=
    \left(\mu_{\rm inj} + 7.85 \times 10^{-11}\right)^{+1.48\times 10^{-9}}_{-1.51\times10^{-9}}.
\end{align}
These tight constraints arise from the large number of inspiral cycles accumulated during the three-month observation in the decihertz band and represent an extraordinary improvement relative to ground-based detector capabilities. Moreover, the onset of dark quadrupole and dipole radiation, governed by $\mu$, lies within the observing frequency band and imprints a characteristic signature on the GW signal, which further leads to precise measurement of $\mu$.

Figure~\ref{fig:gw170817_extrinsic_comparison} shows the marginalized posterior distribution for the extrinsic parameters, namely coalescence time measured in the SSB frame, inclination angle, luminosity distance, ecliptic longitude, and ecliptic latitude for the GW170817-like event. Because \text{IndIGO-D} follows a heliocentric orbit over the three-month observation window, the motion of the detector induces Doppler shifts and time-dependent antenna pattern modulations on the incoming strain. Consequently, the sky position is constrained to exceptional accuracy.
%

%--------------------------------------------------------%
\subsection{Measurement uncertainty on dark photon mass}
\label{subsec:dark_photon_mass_measurement}
%--------------------------------------------------------%
%
A particularly important outcome of measuring the dimensionless parameter $\mu$ is the direct inference of the dark-photon mass, $m_{\rm{A}}$. The two quantities are related by, 
\begin{equation}
% \mu = \frac{G m m_{\rm{A}}}{\hbar c},
m_{\rm{A}} c^2 = \frac{\hbar c^3}{G} \frac{\mu}{m}.
\label{eq:DM_parameters}
\end{equation}
Thus, we can obtain the marginalized posterior for $m_{\rm{A}}$ from posterior samples in $\mu$ and the binary total mass $m$, thereby propagating the correlations and uncertainties in both quantities into the derived $m_{\rm{A}}$ posterior. 
The measurement of the mediator mass is conditional on the detection of a
nonzero dark-sector contribution to the waveform. In the absence of such a
contribution, i.e. in the limit $\alpha,\gamma\rightarrow 0$, the waveform
loses sensitivity to $\mu$ and the dark-photon mass cannot be independently
measured.
For the benchmark GW170817-like system considered here, where the injected
dark-sector parameters produce a detectable modification to the signal, we
obtain
$m_{\rm A}c^2 =
\left(2.42^{+2.08\times10^{-5}}_{-2.09\times10^{-5}}\right)
\times10^{-15}$~eV
at 90\% credibility.
As a numerical validation of the very narrow $\mu$ posterior, we
directly evaluate the likelihood on a fine one-dimensional grid in
$\mu$ across the posterior-support region. The resulting likelihood
peak is smooth and well resolved, with a width consistent with the
strong sensitivity to $\mu$ seen in the Bayesian analysis; when the
remaining parameters are fixed to their injected values, the maximum
occurs at the injected $\mu$.

\begin{figure*}[ht]
  \centering  
  \subfloat[$\alpha=10^{-5}, \, \gamma=0$ \label{fig:alpha1em5_gamma0}]
  {
    \includegraphics[width=0.487\textwidth]{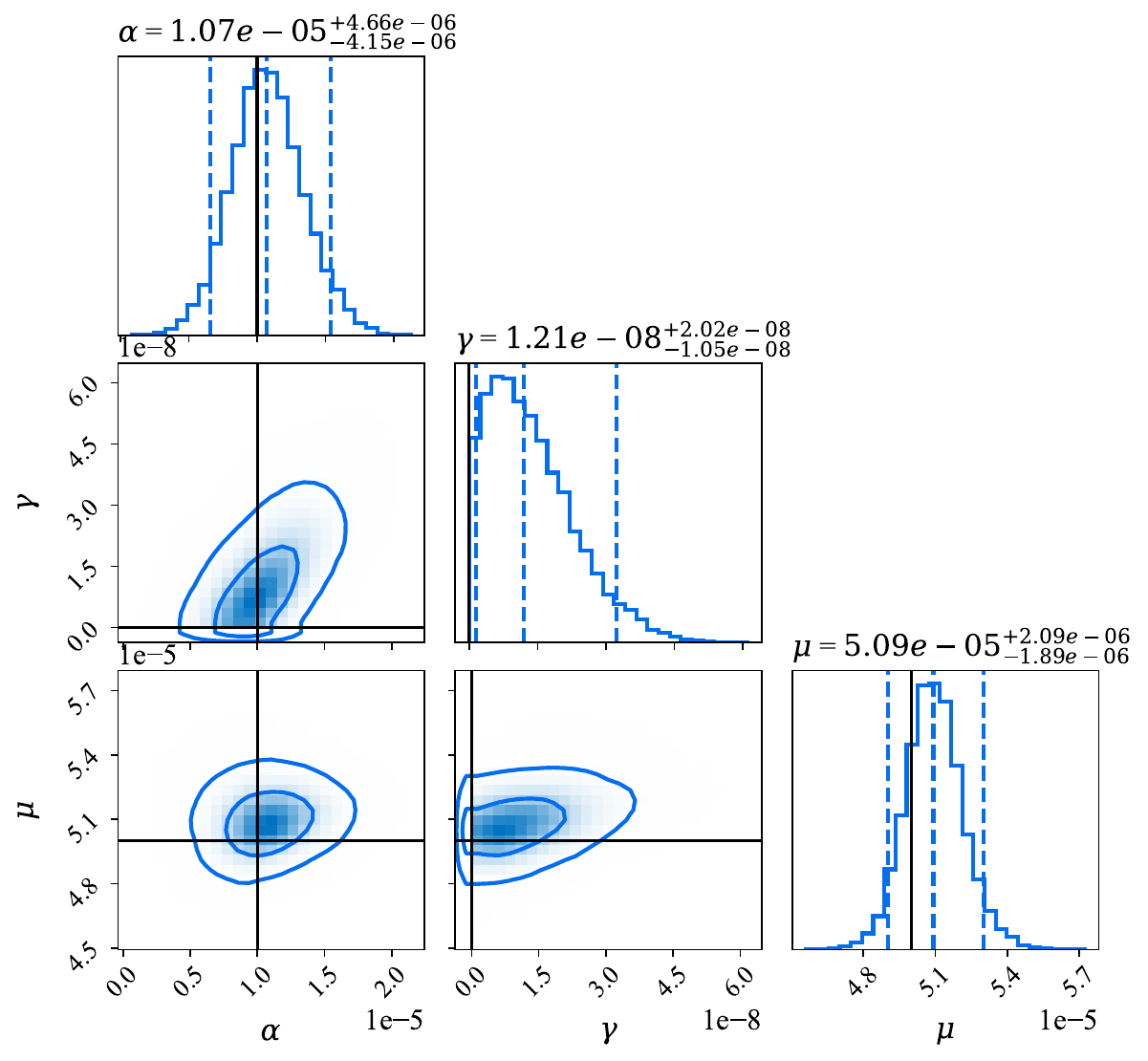}
  }
  \hfill
  \subfloat[$\alpha=10^{-12}, \, \gamma=5\times 10^{-7}$\label{fig:alpha1em12_gamma5em7}]
  {
    \includegraphics[width=0.487\textwidth]{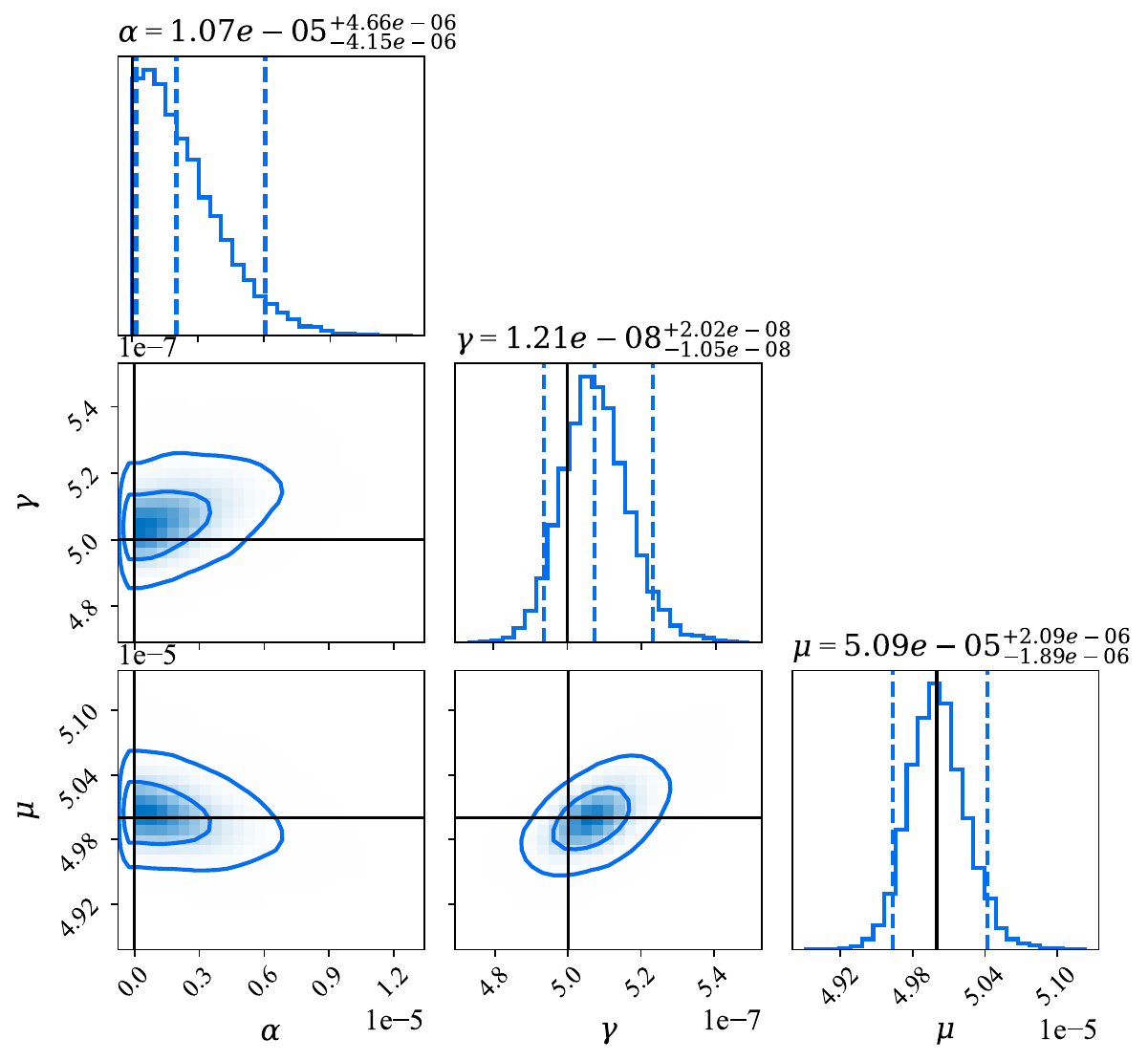}
  }
  %
  % \caption{Marginalized posterior distributions of dark-sector parameters for a GW170817-like BNS system with weaker Yukawa coupling, observed with IndIGO-D. Panel (a) shows the results for $\alpha=10^{-5}$ and $\gamma=0$, for which the injected Yukawa coupling and mediator parameter $\mu$ are recovered with finite uncertainties. Panel (b) shows the results for $\alpha=10^{-12}$ and $\gamma=5\times10^{-7}$, illustrating that although the conservative Yukawa coupling is weak and yields only an upper limit, the non-zero dipole-radiation contribution allows $\gamma$ and $\mu$ to be constrained. The black lines indicate the injected parameter values.}
  \caption{Marginalized posterior distributions of dark-sector parameters for a GW170817-like BNS system with weaker Yukawa coupling, observed with IndIGO-D. Panel (a) shows that both $\alpha$ and $\mu$ can be measured for an injected Yukawa coupling of $\alpha=10^{-5}$ with $\gamma=0$, whereas only an upper limit is obtained on $\gamma$. Panel (b) demonstrates that for $\alpha=10^{-12}$ and $\gamma=5\times10^{-7}$, the conservative Yukawa coupling yields only an upper limit, whereas the non-zero dipole-radiation contribution allows $\gamma$ and $\mu$ to be measured. The black lines indicate the injected parameter values.}
  % \label{fig:two_panel_corner}
\end{figure*}

As a control, we repeat the GW170817-like analysis with a GR injection
($\alpha=\gamma=0$) recovered using the aDM waveform model. For this GR-consistent signal, we instead obtain 90\% upper limits
${\alpha < 3.1\times10^{-4}}$ and
${\gamma < 2.5\times10^{-8}}$, while $\mu$ (and hence the
dark-photon mass), remains unconstrained within the adopted prior.
Thus, for a GR-consistent signal, $\alpha$ and $\gamma$ can instead be bounded, while $\mu$, and hence $m_A$, remains unconstrained within the adopted prior.

%-----------------------------------------------------------%
\subsection{Sensitivity to weaker dark-sector interactions}
\label{subsec:Sensitivity to weaker dark-sector interactions}
%-----------------------------------------------------------%
%
The parameter-estimation analyses presented above adopt benchmark values of the dark-sector parameters motivated by the 90\% credible upper limits obtained by Owen et al.~\cite{Owen:2025odr} from current ground-based GW observations. While these injections demonstrate the measurement precision achievable with IndIGO-D when dark-sector modifications are appreciable, the astrophysical processes responsible for generating and retaining sufficiently large dark charges in NSs impose additional restrictions on the possible interaction strengths.

In particular, Kopp et al.~\cite{Kopp:2018jom} investigated the accumulation of DM in NSs and the conditions under which a long-range dark force could produce observable modifications to compact-binary inspirals. They showed that ordinary DM accumulation during stellar evolution is unlikely to produce sufficiently large net dark charges to yield detectable GW signatures. More recently, Harris and Kahn~\cite{Harris:2025dmg} revisited these constraints by examining different mechanisms through which NSs may acquire dark charge. They demonstrated that the repulsive nature of the long-range dark-vector interaction significantly restricts the amount of charge that can be retained by a NS. For NSs with comparable masses and dark charge, the signal-strength parameter $S$ considered in Ref.~\cite{Harris:2025dmg} is equivalent to the Yukawa coupling strength $\alpha$ in the waveform model considered in our analyses. For DM densities comparable to those in the solar neighborhood, their analysis gives an extremely small characteristic signal strength, ${S \lesssim 10^{-14}}$, while a thermal-production scenario subject to additional fifth-force constraints yields ${S \lesssim 10^{-8}}$. 

Harris and Kahn also discuss the possibility that an exceptionally large dark-matter overdensity near the galactic center could allow NSs to acquire sufficiently large dark charges to produce detectable GW signatures, although the survival of such overdensities is uncertain~\cite{Harris:2025dmg}. 
% Motivated by these astrophysical considerations, we investigate the sensitivity of IndIGO-D to weaker dark-sector interactions through additional injection--recovery analyses.
Motivated by these considerations, we perform additional injection--recovery analyses for the GW170817-like BNS system to investigate the sensitivity of IndIGO-D to weaker dark-sector interactions. We retain the fiducial mediator-mass parameter $\mu = 5 \times 10^{-5}$ and initially set $\gamma = 0$, thereby eliminating the dark dipole-radiation contribution. We first consider an injection with $\alpha = 10^{-8}$, corresponding to the order of magnitude of the signal strength allowed in the thermal-production scenario examined by Harris and Kahn. The resulting posterior distributions do not provide a measurement of the injected Yukawa coupling. Instead, we obtain 90\% credible upper limits $\alpha < 8.82 \times 10^{-6}$ and $\gamma < 4.75 \times 10^{-7}$, while $\mu$ remains unconstrained within its adopted prior range. Thus, for this source configuration, a conservative interaction of this strength does not contain sufficient information in the waveform to determine the mediator mass.

We next investigate the minimum interaction strength at which the Yukawa coupling and mediator mass become simultaneously measurable in the absence of dipole radiation. For a system with $\alpha = 10^{-5}$, $\gamma = 0$, and $\mu = 5 \times 10^{-5}$, we find that both $\alpha$ and $\mu$ are recovered with finite credible intervals containing their injected values, while the dipole parameter is constrained to $\gamma < 2.70 \times 10^{-8}$ with 90\% credibility. Fig.~\ref{fig:alpha1em5_gamma0} shows the corresponding marginalized posterior in the dark-sector parameters. We also performed injections with \(\alpha=10^{-6}\) and \(10^{-7}\), keeping \(\gamma=0\) and \(\mu=5\times10^{-5}\). In both cases, we find that the analysis yields only an upper limit on \(\alpha\), while \(\mu\) remains unconstrained.
% For the systems with $\alpha < 10^{-5}$, we find that only an upper limit on $\alpha$ can be obtained in the absence of dipole radiation.

These results suggest that, for the GW170817-like system and the analysis configuration considered here, the simultaneous measurement of $\alpha$ and $\mu$ becomes possible for Yukawa coupling strengths \(\alpha\gtrsim \mathcal{O}(10^{-5})\), whereas the smaller couplings do not provide sufficient information to determine the mediator mass. This behavior follows from the structure of the waveform corrections; when the leading dipole contribution is absent, the finite-range effects that encode the mediator mass are suppressed as the Yukawa coupling decreases.

Finally, we investigate whether a nonzero dipole-radiation contribution can restore sensitivity to the mediator mass when the conservative Yukawa interaction is negligible. For this purpose, we consider an injection with $\alpha = 10^{-12}$, $\gamma = 5 \times 10^{-7}$, and $\mu = 5 \times 10^{-5}$. In this case, 
the Yukawa coupling remains consistent with zero and only an upper limit on $\alpha$ is obtained. However, both $\gamma$ and $\mu$ are recovered with finite credible intervals as shown in Fig.~\ref{fig:alpha1em12_gamma5em7}. 

This result demonstrates that the measurement of the mediator mass need not rely on a measurable conservative Yukawa interaction.
The mediator mass may remain measurable if the dissipative dark-sector contribution is sufficiently large even when the Yukawa coupling is too small to be resolved. Taken together, these additional analyses illustrate the different roles of conservative and dissipative dark-sector modifications in determining the measurability of $\mu$.
%-----------------------------------------------------------%
\subsection{Timing accuracy in the SSB and detector frames}
\label{subsec:timing_vertex}
%-----------------------------------------------------------%
%
\begin{figure}[t!]
    \centering
    \includegraphics[width=\linewidth]{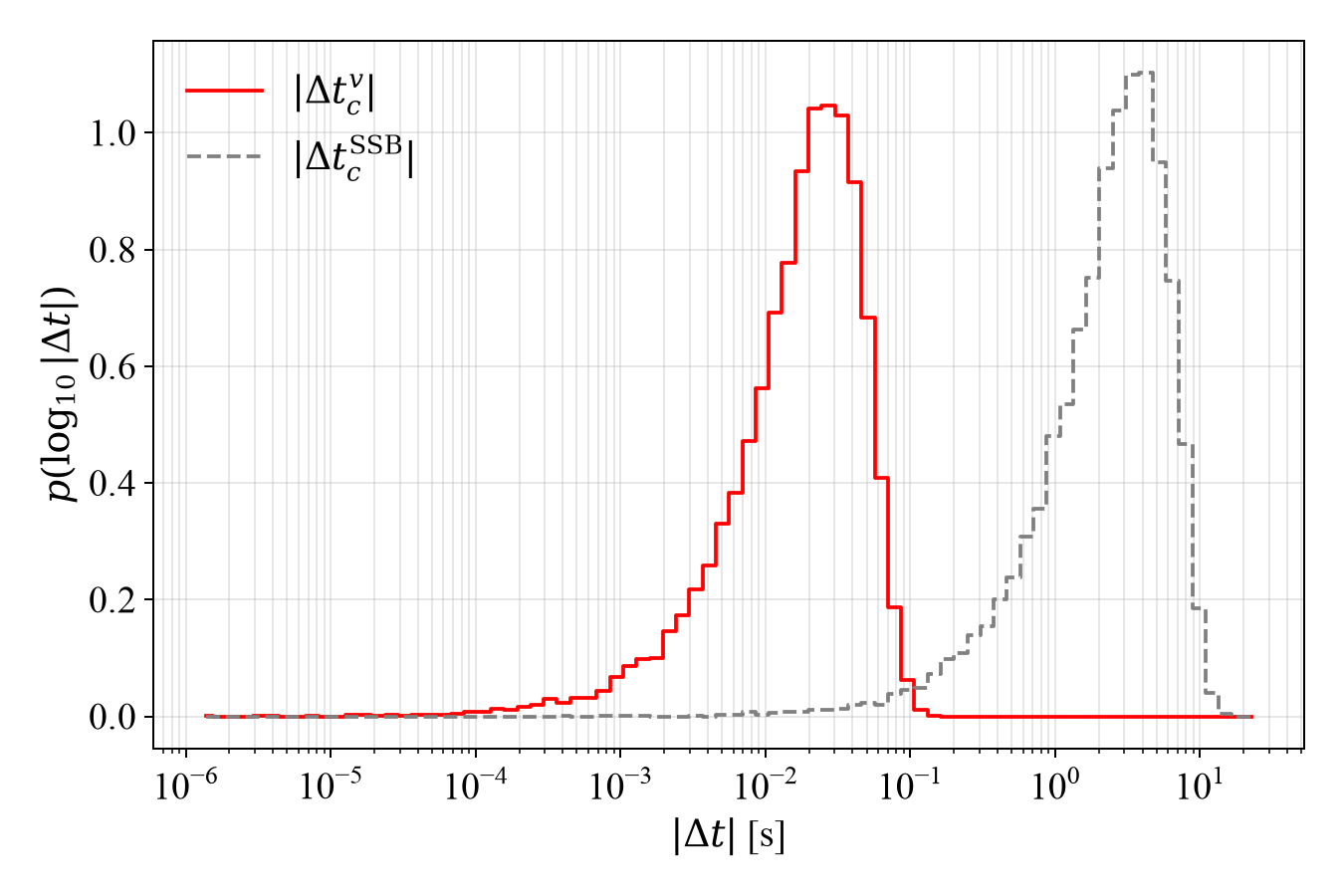}
    \caption{Marginalized posterior distributions of the timing offsets
    $|\Delta t_c^{\rm SSB}| = |t_c^{\rm SSB} - t_{c,{\rm inj}}^{\rm SSB}|$
    and
    $|\Delta t_c^{v}| = |t_c^{v} - t_{c,{\rm inj}}^{v}|$
    for the GW170817-like binary in IndIGO-D event also shown in Figure~\ref{fig:gw170817_extrinsic_comparison}. The detector-frame coalescence time at the Michelson vertex, $t_c^v$, is much more tightly constrained than the SSB-frame coalescence time $t_c^{\rm SSB}$, because the latter is broadened by the sky-dependent propagation delay between the SSB and the detector.}
    \label{fig:tc_vertex_vs_ssb}
\end{figure}
It is useful to distinguish between the coalescence time referred to the solar-system barycenter (SSB), $t_c^{\rm SSB}$, and the coalescence time referred to the detector itself. In the present IndIGO-D configuration, the most natural detector-frame time is the coalescence time at the central Michelson vertex spacecraft, which we denote by $t_c^v$. This distinction follows directly from the detector response written in Eq.~(\ref{eq:detector_response}), where the phase contains the detector-position term $\hat{k}\cdot \vec{x}_d(t)/c$, and from the time-frequency map of Eq.~(\ref{eq:time_frequency}), which evaluates the detector response at the epoch corresponding to each instantaneous GW frequency.

Because the measured signal is received at the moving detector rather than at the SSB, the two coalescence times are related by the light-travel delay between the SSB and the vertex spacecraft,
\begin{equation}
    t_c^v = t_c^{\rm SSB} + \frac{\hat{k}\cdot \vec{x}_v(t_c^v)}{c},
    \label{eq:tc_vertex_relation}
\end{equation}
where $\vec{x}_v(t)$ is the SSB position of the vertex spacecraft and $\hat{k}$ is the propagation direction of the wave. Equation~(\ref{eq:tc_vertex_relation}) shows that $t_c^{\rm SSB}$ contains an additional sky-dependent timing contribution through the detector position, whereas $t_c^v$ is the time more directly associated with the physical arrival of the signal at the detector.

As a consequence, the posterior for $t_c^{\rm SSB}$ is broadened by its correlation with the sky-location parameters $(\lambda,\beta)$, while the detector-frame time $t_c^v$ is significantly better constrained. Figure~\ref{fig:tc_vertex_vs_ssb} illustrates this effect through the marginalized posterior distributions of 
$|\Delta t_c^{\rm SSB}| = |t_c^{\rm SSB} - t_{c,{\rm inj}}^{\rm SSB}|$
and
$|\Delta t_c^{v}| = |t_c^{v} - t_{c,{\rm inj}}^{v}|$ corresponding to zero-noise analysis of GW170817-like event. The 90\% credible width decreases from $12.16~{\rm s}$ for $t_c^{\rm SSB}$ to $0.097~{\rm s}$ for $t_c^v$, corresponding to an improvement by a factor of $\sim 124$. Similarly, for the noisy injection, the 90\% credible widths are $12.2$~s and $0.11$~s for $t_c$ measured at SSB and the detector vertex, respectively with an improvement by a factor of $\sim 110$.
%The same behavior is also seen in the standard deviation and 68\% credible width.

This demonstrates that, for a heliocentric detector such as IndIGO-D, the detector-frame coalescence time is the more sharply measured timing parameter, while the SSB-frame coalescence time inherits an additional uncertainty from the sky-dependent barycentric delay.

%%================================%%
\section{Summary and Conclusions}
\label{sec:conclusion}
%%================================%%
%
In this work, we have investigated the prospects of probing aDM interactions using long-duration GW observations of compact binary inspirals in the decihertz frequency band. We considered neutron stars carrying an effective dark charge under a massive dark $U(1)_D$ vector field. 
The resulting interaction modifies both the conservative binary dynamics through a Yukawa potential and the dissipative dynamics through dark-photon radiation.
The corresponding modifications to the GW amplitude and phase are parameterized by the Yukawa coupling strength $\alpha$, the dipole-radiation parameter $\gamma$, and the dimensionless dark-photon mass parameter
$\mu$.

An important feature of the massive-vector model is the kinematic activation of the dark quadrupole and dipole radiation channels at the characteristic frequencies $f_{\rm quad}$ and $f_{\rm dip}$, respectively. For current ground-based observations these frequencies generally lie below the sensitive detector band for the range of $\mu$ in which the waveform model is valid. In the decihertz band, however, the thresholds can lie directly within the observed inspiral. The waveform must therefore be treated carefully when the binary evolves across these activation frequencies. The activation of dark-radiation channels require the inspiral phase to be evaluated piecewise across $f_{\rm quad}$ and $f_{\rm dip}$. We ensure continuity of the Fourier-domain phase and its first frequency derivative by introducing appropriate matching constants, thereby removing spurious phase discontinuities. The Fourier amplitude, however, retains finite jumps associated with the step-function approximation to the radiation flux.

%In particular, the terms involving Heaviside function appearing in the radiated power cause the analytic expression for $E'(x)/P(x)$ to change across the two thresholds and introduce unphysical discontinuities in the Fourier-domain phase. 

% In particular, the terms involving Heaviside functions
% appearing in the radiated power cause the analytic
% expression for $E'(x)/\mathcal{P}(x)$ to change across the two
% thresholds. If the corresponding phase expressions are evaluated
% piecewise without restoring the appropriate integration constants,
% spurious discontinuities appear in the Fourier-domain phase.
% Therefore, we divide the inspiral into three regions: one in which both dark-radiation channels are closed, followed by the one in which only the quadrupole channel is open, and finally, one in which both channels are open. In each region, we supplement the stationary-phase expression by the matching terms where the constants $A_k$ and $B_k$ are fixed by requiring continuity of the phase and its derivative with respect to frequency across $f_{\rm quad}$ and $f_{\rm dip}$. The region containing coalescence is used as the reference, with $A_3=B_3=0$. This prescription restores the integration constants that are lost when the phase is evaluated piecewise and removes otherwise spurious phase jumps. The Fourier amplitude, in contrast, is a pointwise function of the instantaneous energy-loss rate and therefore retains the finite jumps associated with the step-function treatment of the opening radiation channels.

Before turning to decihertz observations, we also translated existing LVK parameterized-GR bounds into approximate constraints on the aDM parameters. In the ground-based frequency band, where both dark-radiation channels are already open, the aDM corrections reduce to modifications of the usual PN coefficients. The $-1$PN bounds from GW170817 and GW230529 yield $\gamma \lesssim 1.2\times10^{-4}$ and $\gamma \lesssim 1.3\times10^{-3}$, respectively, while the 1PN bound from GW170817 gives $\alpha\lesssim10^{-1}$. However, existing parameterized tests do not include the deformation corresponding to the leading $-2$PN aDM contribution, proportional to $\alpha\mu^2$, limiting their ability to constrain the mediator-mass parameter $\mu$. This provides additional motivation for observations at lower frequencies, where the contribution of the negative-PN corrections will be more pronounced and may lead to more precise estimates of the dark-sector parameters.

We quantify this prospect by carrying out full Bayesian parameter-estimation analyses for representative BNS and NSBH systems observed with the proposed IndIGO-D decihertz detector. We simulate BNS systems resembling GW170817 and GW190425, and NSBH systems resembling GW200105, GW200115, and GW230529, incorporating the aDM corrections into the \texttt{IMRPhenomD\_NRTidalv2} and \texttt{IMRPhenomNSBH} waveform models, respectively. Considering an in-band duration of three months in the IndIGO-D observatory, direct likelihood evaluation is computationally expensive, and we therefore employed the relative binning approximation to accelerate inference.

To guide the selection of suitable prior ranges, we examined two-dimensional match surfaces in the intrinsic parameter space and adopted ranges broader than the local high-match regions to encompass the expected posterior support while maintaining the accuracy of relative binning. The Bayesian analyses demonstrate the sensitivity of decihertz observations to the aDM parameters. For the GW170817-like BNS system, the zero-noise analysis constrains $\alpha$, $\gamma$, and $\mu$ with $90\%$ credible uncertainties of approximately $3.0\times10^{-5}$, $1.4\times10^{-7}$, and $1.5\times10^{-9}$, respectively. An analysis including stationary Gaussian noise yields comparable posterior widths and correlations, demonstrating that the inference is robust against a representative detector-noise realization. Similar measurements are obtained for the GW190425-like BNS system. The NSBH systems, with $\alpha=0$, also yield informative constraints on $\gamma$ and $\mu$. These results highlight the sensitivity of the decihertz band to dark-sector interactions, owing to the large number of accumulated inspiral cycles and the enhanced contributions of negative-PN corrections at low frequencies.

A particularly significant result is the strong sensitivity to the dark-photon mass when the dark-sector interaction is nonzero and the activation frequencies lie within the observing band.
For the GW170817-like benchmark system, we obtain
$m_A c^2 =
\left(2.42^{+2.08\times10^{-5}}_{-2.09\times10^{-5}}\right)
\times10^{-15}\ {\rm eV}$
at $90\%$ credibility, corresponding to an absolute uncertainty of order $10^{-20}\,{\rm eV}$.
This measurement is conditional on a detectable aDM contribution to the waveform and on the benchmark nonzero dark-sector couplings adopted for the injection.
For a GR-consistent signal, we instead obtain upper limits on $\alpha$
and $\gamma$, while $\mu$, and hence the dark-photon mass, remains
unconstrained within the adopted prior.
By contrast, for the benchmark aDM signal considered here, IndIGO-D
constrains $\mu$ sharply, whereas present ground-based observations
leave $\mu$ effectively unconstrained beyond the validity prior of the
waveform model.

We further investigated the sensitivity of IndIGO-D to weaker dark-sector interactions, motivated by astrophysical considerations that restrict the accumulation of large net dark charges in NSs~\cite{Kopp:2018jom,Harris:2025dmg}. For the GW170817-like system and analysis configuration considered here, Yukawa interactions at the level of $\alpha \sim 10^{-8}$ do not permit an informative mediator-mass measurement when $\gamma = 0$, while Yukawa couplings of order $\alpha \sim 10^{-5}$ allow both $\alpha$ and $\mu$ to be recovered. We further find that a nonzero charge-to-mass asymmetry between the NSs can provide sufficient information to constrain both $\gamma$ and $\mu$, even when the Yukawa coupling is negligible ($\alpha=10^{-12}$). 
% These results demonstrate that the measurability of the mediator mass depends on the strength of the dark-sector modifications and that information about $\mu$ can arise from either the conservative or dissipative sector. The corresponding astrophysical viability remains dependent on the mechanisms through which NSs acquire their individual dark charges.
Thus, information about $\mu$ can arise through either the conservative Yukawa sector or the dissipative dipole-radiation sector. The astrophysical viability of these signals remains dependent on the mechanisms through which NSs acquire and retain their individual dark charges.

These results demonstrate that the decihertz frequency band provides a particularly favorable regime for probing long-range dark-sector interactions. The prolonged early inspiral enhances negative-PN modifications, while the possibility of directly observing the onset of massive dark-radiation channels introduces additional information about the mediator mass. Future decihertz GW observatories could therefore provide not only stringent constraints on effective dark charges and dipole radiation, but also direct measurements of the mass scale and interaction range of an ultralight dark mediator. More generally, these results illustrate the broader potential of long-duration compact-binary inspirals as precision laboratories for fundamental physics beyond general relativity and the Standard Model, and it will be interesting to extend these studies to other effective field theory models of DM (e.g., \cite{Miller:2025yyx, Bhattacharyya:2023kbh}).

\begin{acknowledgments}
We are grateful to the HPC facility at IIT Gandhinagar for access to computational resources, and especially to Raviraj Sukhadiya for reliable support in maintaining the cluster. D.~Tahelyani acknowledges the support from the Inspire fellowship from the Department of Science and Technology of India, under fellowship number DST/INSPIRE Fellowship/[IF210643]. A.~Sharma acknowledges IUCAA for the postdoctoral fellowship. 
We thank our colleagues in the IndIGO consortium for suggestions and helpful discussions and providing the IndIGO-D noise curve used in this work.
\noindent
We acknowledge the National Supercomputing Mission (NSM) for providing computing resources of ‘PARAM Ananta’ at IIT Gandhinagar, which is implemented by C-DAC and supported by the Ministry of Electronics and Information Technology (MeitY) and Dept. of Science and Technology (DST), Govt. of India. AB is supported by the Core Research Grant (CRG/2023/005112) by ANRF (erstwhile SERB) of the Indian Government. We acknowledge the use of the ChatGPT (OpenAI, GPT-5) to improve the scientific language and clarity of the manuscript. \newline
\textit{Softwares.} We used the \textsc{PyCBC} library~\cite{pycbc_software} for waveform generation and likelihood evaluation, implementing the dark-sector corrections through a custom waveform plugin. Bayesian parameter estimation was performed using \textsc{Bilby}~\cite{Ashton:2018jfp} with the \texttt{dynesty} nested sampler~\cite{Speagle:2019ivv}. Numerical computations and visualization employed \textsc{NumPy}~\cite{Harris:2020xlr}, \textsc{SciPy}~\cite{Virtanen:2019joe}, and \textsc{Matplotlib}~\cite{Hunter:2007ouj}.
\end{acknowledgments}
%\clearpage
\appendix

\begin{widetext}

%%===========================================%%
\section{Explicit post-Newtonian coefficients}
\label{app:coefficients}
%%===========================================%%
%
%---------------------------------------------%
\subsection{Binding energy and radiated power}
\label{app:energy_flux}
%---------------------------------------------%
%
For completeness, we give here the explicit expressions for the conserved binding energy and total radiated power used in the waveform construction. To the order retained in this work~\cite{Owen:2025odr}, the binding energy is
\begin{equation}
E(x) = -\frac{1}{2}\eta m c^2 x 
        \bigg\{1-\alpha\Big(\frac{2}{3}-2\frac{\mu}{x}+\frac{5}{3}\frac{\mu^2}{x^2}\Big) 
        -x\Big(\frac{3}{4}+\frac{\eta}{12}\Big)
        +\alpha x\Big[\frac{31}{9}\eta-\frac{2}{3}\eta_\gamma\frac{\mu}{x}\Big] 
        -\Big[\frac{5}{3}(1+3\eta)+(4+6\eta)\ln\frac{\mu}{x}\Big]\frac{\mu^2}{x^2}\bigg\}\, ,
\label{eq:Ex_closed}
\end{equation}
where $\eta_\gamma\equiv1+\eta+(2-4\eta)\gamma_E$ and $\gamma_E$ is the Euler--Mascheroni constant. The absence of threshold factors reflects the conservative nature of $E(x)$.

The corresponding total radiated power is
%\begin{widetext}
    \begin{align}
\mathcal{P}(x) = &\frac{c^5}{G}\frac{32}{5}\eta^2 x^5\bigg\{1-\frac{4}{3}\alpha\Big(1-\frac{1}{2}\frac{\mu^2}{x^2}\Big)-x\Big(\frac{1247}{336}-\frac{35}{12}\eta\Big) +\alpha x\Big[\frac{1739}{504}-\frac{29}{18}\eta-\frac{4}{3}\eta_\gamma\frac{\mu}{x}+\Big(\frac{1933}{1008}-\frac{197}{84}\eta\Big)\frac{\mu^2}{x^2}\Big] \nonumber\\
&+\frac{5}{32}\Big[\frac{2}{3}\frac{\gamma}{x}-2\alpha\Delta-\frac{\gamma}{9}(3+9\Delta+2\eta)\Big]\Theta\Big(\frac{x^{3/2}}{\mu}-1\Big) +\frac{5}{32}\Big[-\frac{8}{5}\alpha(\Delta-1)+\frac{2}{5}\gamma(1-\Delta)^2\Big]\Theta\Big(\frac{x^{3/2}}{\mu}-\frac{1}{2}\Big)\bigg\}\, ,
\label{eq:Px_closed}
\end{align}
%\end{widetext}
where $\Delta\equiv(m_1-m_2)/m$. Unlike the binding energy, $\mathcal{P}(x)$ contains the Heaviside functions that mark the kinematic activation of the dipole and quadrupole dark-photon radiation channels.

%--------------------------------------------%
\subsection{Amplitude and phase coefficients}
%--------------------------------------------%

This appendix collects the explicit closed-form coefficients $\tilde A_{(n)}(x)$ and $\tilde\varphi_{(n)}(x)$, $n=-2,-1,0,1$, entering the amplitude and phase series of Eq.~\eqref{eq:h_series} (Sec.~\ref{subsec:gravitational_waveform}). They are obtained by substituting the closed-form energy $E(x)$ and flux $\mathcal{P}(x)$ (Eqs.~\eqref{eq:Ex_closed}--\eqref{eq:Px_closed}) into the general SPA relations of Eq.~\eqref{eq:t0_phi_omegadot}, expanding to linear order in $\alpha,\gamma$ and to $\mathcal{O}(\mu^2/x^2)$, and collecting the result as a series in $x$, following Ref.~\cite{Owen:2025odr}.
%
%\enlargethispage{2\baselineskip}
%\newpage
%\begin{widetext}
\begin{subequations}
\label{eq:adm_amp_phase}
\begin{align}
\tilde{A}_{(-2)} &= \alpha \mu^2 \frac{5}{6} \, ,
\label{eq:adm_Am2} \\
\tilde{A}_{(-1)} &= \alpha \mu^2 \left( \frac{1051}{576} + \frac{301}{48}\eta \right)
    - \gamma \frac{5}{96} \Theta \left( \frac{x^{3/2}}{\mu} - 1 \right) \, ,
\label{eq:adm_Am1} \\
\tilde{A}_{(0)} &= 1 - \alpha \left[ \frac{1}{3}(1 + \eta_\gamma \mu)
    - \frac{5}{32} \Delta \Theta \left( \frac{x^{3/2}}{\mu} - 1 \right)
    + \frac{1}{8}(1 - \Delta) \Theta \left( \frac{x^{3/2}}{\mu} - \frac{1}{2} \right) \right] \nonumber \\
    &\quad - \gamma \left[ \left( \frac{5945}{64512} + \frac{1475}{5376}\eta
    - \frac{5}{64}\Delta \right) \Theta \left( \frac{x^{3/2}}{\mu} - 1 \right)
    + \frac{1}{32}(1 - \Delta)^2 \Theta \left( \frac{x^{3/2}}{\mu} - \frac{1}{2} \right) \right] \, ,
\label{eq:adm_A0} \\
\tilde{A}_{(1)} &= -\left( \frac{323}{224} - \frac{451}{168}\eta \right)
    + \alpha \left( \frac{4895}{2016} - \frac{485}{504}\eta \right) \, ,
\label{eq:adm_A1} \\
\tilde{\varphi}_{(-2)} &= \alpha \mu^2 \frac{10}{27} \, ,
\label{eq:adm_phim2} \\
\tilde{\varphi}_{(-1)} &= \alpha \mu^2 \left( \frac{311}{126} + \frac{264}{49}\eta \right)
    - \gamma \frac{5}{84} \Theta \left( \frac{x^{3/2}}{\mu} - 1 \right) \, ,
\label{eq:adm_phim1} \\
\tilde{\varphi}_{(0)} &= 1 + \alpha \left[ \frac{2}{3}(1 + \eta_\gamma \mu)
    + \frac{5}{16} \Delta \Theta \left( \frac{x^{3/2}}{\mu} - 1 \right)
    - \frac{1}{4}(1 - \Delta) \Theta \left( \frac{x^{3/2}}{\mu} - \frac{1}{2} \right) \right] \nonumber \\
    &\quad - \gamma \left[ \left( \frac{4555}{8064} + \frac{5}{9}\eta
    - \frac{5}{32}\Delta \right) \Theta \left( \frac{x^{3/2}}{\mu} - 1 \right)
    + \frac{1}{16}(1 - \Delta)^2 \Theta \left( \frac{x^{3/2}}{\mu} - \frac{1}{2} \right) \right] \, ,
\label{eq:adm_phi0} \\
\tilde{\varphi}_{(1)} &= \left( \frac{3715}{756} + \frac{55}{9}\eta \right)
    + \alpha \left( \frac{355}{81} + \frac{2540}{81}\eta \right) \, .
\label{eq:adm_phi1}
\end{align}
\end{subequations}
%\end{widetext}
Setting both $\alpha$ and $\gamma$ to zero in Eq.~\eqref{eq:adm_amp_phase} recovers the Newtonian and 1PN contributions to the amplitude and phase of the {\IMRPhenomD} waveform family~\cite{Khan:2015gaa, Husa:2015iqa}, as required.

\end{widetext}

\bibliography{references}
\end{document}